\documentclass[reprint,twocolumn]{revtex4-2}
\usepackage{amsfonts}
\usepackage{amsmath}
\usepackage{amssymb}
\usepackage{mathtools}
\usepackage{charter}
\usepackage{graphicx}
\usepackage{float}
\usepackage{amsmath}
\usepackage{amssymb}
\usepackage{amsfonts}
\usepackage{charter}
\usepackage{graphicx}
\usepackage{subfigure}
\usepackage{graphicx}
\usepackage{epstopdf}
\usepackage{color}
\usepackage{tocvsec2}
\usepackage{enumerate}
\usepackage{graphicx}
\usepackage{dcolumn}
\usepackage{bm}
\usepackage{romannum}
\usepackage{fancyhdr}
\usepackage[hidelinks]{hyperref}

\begin{document}

\title{Quantum Steering and Entanglement in a Tritter: Hierarchy under Loss}
\author{Jifeng Sun$^{1,2}$}
\author{Shumin Yang$^{1}$}
\author{Teng Zhao$^{2,3}$}
\author{Qingqian Kang$^{2,4}$}
\author{Liyun Hu$^{2,3}$}
\email{hlyun@jxnu.edu.cn}
\affiliation{$^{1}$College of Electronic and Information Engineering, Nanchang Institute
of Technology, Nanchang 330044, China}
\affiliation{$^{2}$Center for Quantum Science and Technology, Jiangxi Normal University,
Nanchang 330022, China}
\affiliation{$^{3}$Institute for Military-Civilian Integration of Jiangxi Province,
Nanchang 330200, China}
\affiliation{$^{4}$School of Photoelectric Engineering, Jiangxi Modern Polytechnic College, Nanchang 330022, China}

\begin{abstract}
We present a comprehensive phase diagram of quantum correlations in a
three-mode Gaussian state generated by a tritter, driven by a two-mode
squeezed vacuum and a coherent state. The coherent amplitude does not affect
the correlation structure, which is solely governed by the initial
squeezing. By systematically analyzing five physically relevant asymmetric
loss configurations, we map out the exact resilience thresholds for
entanglement and Einstein-Podolsky-Rosen (EPR) steering under all bipartite
partitions. We reveal that EPR steering exhibits a pronounced directional
asymmetry under loss, and its survival can be maintained over a much wider
range of loss by strategically protecting a single channel. This tunable
fragility provides practical guidance for one-sided device-independent
quantum protocols in noisy asymmetric networks. We further confirm the
limitations of R\'{e}nyi-2 entropy in the quantification of entanglement and
steering. Our results transform the abstract correlation hierarchy into a
calculable, experimentally relevant guide for engineering robust quantum
resources.
\end{abstract}

\keywords{Quantum steering, Continuous-variable entanglement, Tritter,
Gaussian states, Loss channels}
\maketitle

\section{Introduction}

Quantum entanglement stands as a cornerstone of quantum mechanics,
epitomizing nonlocal correlations absent in classical physics \cite%
{Einstein1935, Schrodinger1935a, Schrodinger1936}. In continuous-variable
(CV) systems, Gaussian states---those with Gaussian Wigner
functions---provide a fertile testbed for exploring multipartite
entanglement due to their experimental accessibility and theoretical
tractability \cite{Weedbrook2012, Adesso2007}. Beyond entanglement,
Einstein-Podolsky-Rosen (EPR) steering captures a distinctive asymmetric
form of quantum correlation where one party can, by local measurements,
``steer'' or affect the state of a distant party \cite{Wiseman2007,
Jones2007, Cavalcanti2007}. This asymmetry grants steering unique advantages
in applications such as one-sided device-independent quantum key
distribution \cite{Branciard2012, Zhou2020}. The resource theory of quantum
steering has been developed to quantify and understand its applications and
limitations \cite{Li2023,Li2025,Yang2026}.

The generation of multipartite entangled states often relies on linear
optical networks \cite{Wang2022}. Among these, the tritter---a three-port
balanced beam splitter---is a fundamental element for generating and
manipulating three-mode Gaussian states \cite{Spagnolo2013, Ferraro2014}.
When fed with a two-mode squeezed vacuum (TMSV) state and a coherent state,
the tritter output constitutes a rich resource of tripartite quantum
correlations \cite{Chang2022}. While the hierarchy between entanglement and
steering has been studied in various contexts \cite%
{Kogias2015,Yu2017,Adesso2012}, a systematic analysis of this hierarchy in a
tritter-generated three-mode Gaussian state, particularly under realistic
loss conditions with various asymmetric configurations, remains lacking.
Such an analysis is crucial for practical quantum communication where
different channels often exhibit different loss characteristics, and it is
essential to provide quantitative operational guidance for optimizing
resources.

Recent studies have advanced our understanding of steering in lossy
environments and multipartite settings. The quantification of steering under
attenuation has been addressed in continuous-variable systems, providing
general bounds for loss tolerance \cite{PRXQuantum2022}. Multipartite
steering in lossy networks has been explored, highlighting the importance of
asymmetric channel configurations \cite{PRX2022}. The entanglement structure
of multimode Gaussian states has been characterized, offering tools to
analyze complex correlation patterns \cite{PRResearch2020}. These works
underline the need for a detailed, analytically tractable platform where the
interplay between loss distribution and directional steerability can be
mapped out exactly---a gap that our work fills.

In this work, we go beyond merely observing this hierarchy. We provide a
comprehensive and quantitative phase diagram of quantum correlations in a
realistic tritter network. Specifically, we address the practical question:
given a certain amount of loss, how does its distribution across the
channels affect the survival of entanglement and, more critically,
directional steerability? To this end, we introduce a systematic model of
five distinct asymmetric loss configurations, which are representative of
realistic network scenarios. By deriving exact analytical criteria for
entanglement and EPR steering under all possible partitions, we map out
their resilience thresholds. This analysis yields actionable insights,
demonstrating, for instance, that steering in one direction can be
maintained over a much wider range of loss by strategically protecting a
single channel, a finding with direct implications for resource allocation
in quantum networks. Our work transforms the abstract hierarchy of quantum
correlations into a concrete, calculable operational guide for a key CV
multipartite platform.

Furthermore, We have examine the performance of the R\'{e}nyi-2
entropy-quantified multi-system entanglement and steering in this paper and
found that it is not applicable to the scenarios discussed here. We also
discuss potential extensions to $m$-mode networks, providing insights into
the scalability of our findings.

The remainder of this paper is organized as follows. In Sec.~\ref{sec:model}%
, we introduce the tritter setup and derive the covariance matrix of the
generated three-mode Gaussian state. Section~\ref{sec:ideal_correlations}
presents a detailed analysis of the entanglement and EPR steering
hierarchies under ideal, lossless conditions, including all bipartite and
tripartite partitions. In Sec.~\ref{sec:loss}, we model various
configurations of asymmetric and symmetric loss channels and investigate
their distinct impacts on both types of quantum correlations. We
systematically compare their robustness and explicitly demonstrate the
strict inclusion relation. Section~\ref{sec:discussion} discusses the
implications, connection to other measures, the generality of our findings,
and an outlook for larger systems. Finally, Sec.~\ref{sec:conclusion}
summarizes our main findings.

\section{Theoretical Framework}

\label{sec:model}

\subsection{The Tritter Operation and Output State}

The tritter is a three-port linear optical device as shown in Fig.~\ref%
{fig:tritter}, which can be described by a unitary matrix $U$ that
transforms the input annihilation operators $\hat{a}, \hat{b}, \hat{c}$ to
the output operators $\hat{a}_1, \hat{b}_1, \hat{c}_1$ \cite{Spagnolo2013}:
\begin{equation}
\begin{pmatrix}
\hat{a}_1 \\
\hat{b}_1 \\
\hat{c}_1%
\end{pmatrix}
= U
\begin{pmatrix}
\hat{a} \\
\hat{b} \\
\hat{c}%
\end{pmatrix}%
, \quad \text{with } U = \frac{1}{\sqrt{3}}
\begin{pmatrix}
1 & e^{2i\pi/3} & e^{2i\pi/3} \\
e^{2i\pi/3} & 1 & e^{2i\pi/3} \\
e^{2i\pi/3} & e^{2i\pi/3} & 1%
\end{pmatrix}%
.  \label{eq:tritter_U}
\end{equation}

We consider the input state to be a product of a two-mode squeezed vacuum
(TMSV) between modes $a$ and $b$, and a coherent state $|\gamma\rangle_c$ in
mode $c$:
\begin{equation}
|\psi_{\text{in}}\rangle = |\text{TMSV}(r)\rangle_{ab} \otimes
|\gamma\rangle_c,  \label{eq:input_state}
\end{equation}
where $|\text{TMSV}(r)\rangle_{ab} = \frac{1}{\cosh r}\sum_{n=0}^{\infty}
(\tanh r)^n |n,n\rangle_{ab}$, with $r \geq 0$ being the squeezing
parameter, and $|\gamma\rangle_c = e^{\gamma \hat{c}^\dagger - \gamma^*\hat{c%
}}|0\rangle_c$.

The squeezing operator is defined as $\widehat{S}(r) = \exp[r(\hat{a}\hat{b}
- \hat{a}^\dagger\hat{b}^\dagger)]$, and the displacement operator as $%
\widehat{D}_c(\gamma) = \exp(\gamma \hat{c}^\dagger - \gamma^* \hat{c})$.
The explicit Bogoliubov transformation of the tritter in the symplectic
representation is given by a $6\times6$ matrix $S$, which relates the output
quadrature vector to the input one via $\hat{\bm{\xi}}_{\mathrm{out}}=S\hat{%
\bm{\xi}}_{\mathrm{in}}$. The transformation $S$ can be constructed from the
unitary matrix $U$ by mapping $U\to S$ according to the standard embedding
of unitary into symplectic matrices. The full form is provided in the
Supplemental Material. The output state after the tritter operation can be
expressed as:
\begin{equation}
|\psi_{\text{out}}\rangle = \frac{1}{\cosh r} \exp\left(-\frac{|\gamma|^2}{2}
+ \widehat{O}_{T1} + \widehat{O}_{T2}\right) |000\rangle_{abc},
\label{eq:output_state_explicit}
\end{equation}
where $\widehat{O}_{T1} = \frac{\lambda}{3}\left[e^{-2i\pi/3}(\hat{a}%
^{\dagger 2} + \hat{b}^{\dagger 2}) + e^{2i\pi/3}\hat{c}^{\dagger 2} +
e^{i\pi/3}\hat{a}^\dagger\hat{b}^\dagger\right]$ and $\widehat{O}_{T2} =
\frac{\gamma}{\sqrt{3}}\left[e^{-2i\pi/3}(\hat{a}^\dagger + \hat{b}^\dagger)
- \hat{b}^\dagger\hat{c}^\dagger - \hat{a}^\dagger\hat{c}^\dagger\right]$,
with $\lambda = \tanh r$.

\begin{figure}[tbh]
\centering
\includegraphics[width=0.8\columnwidth]{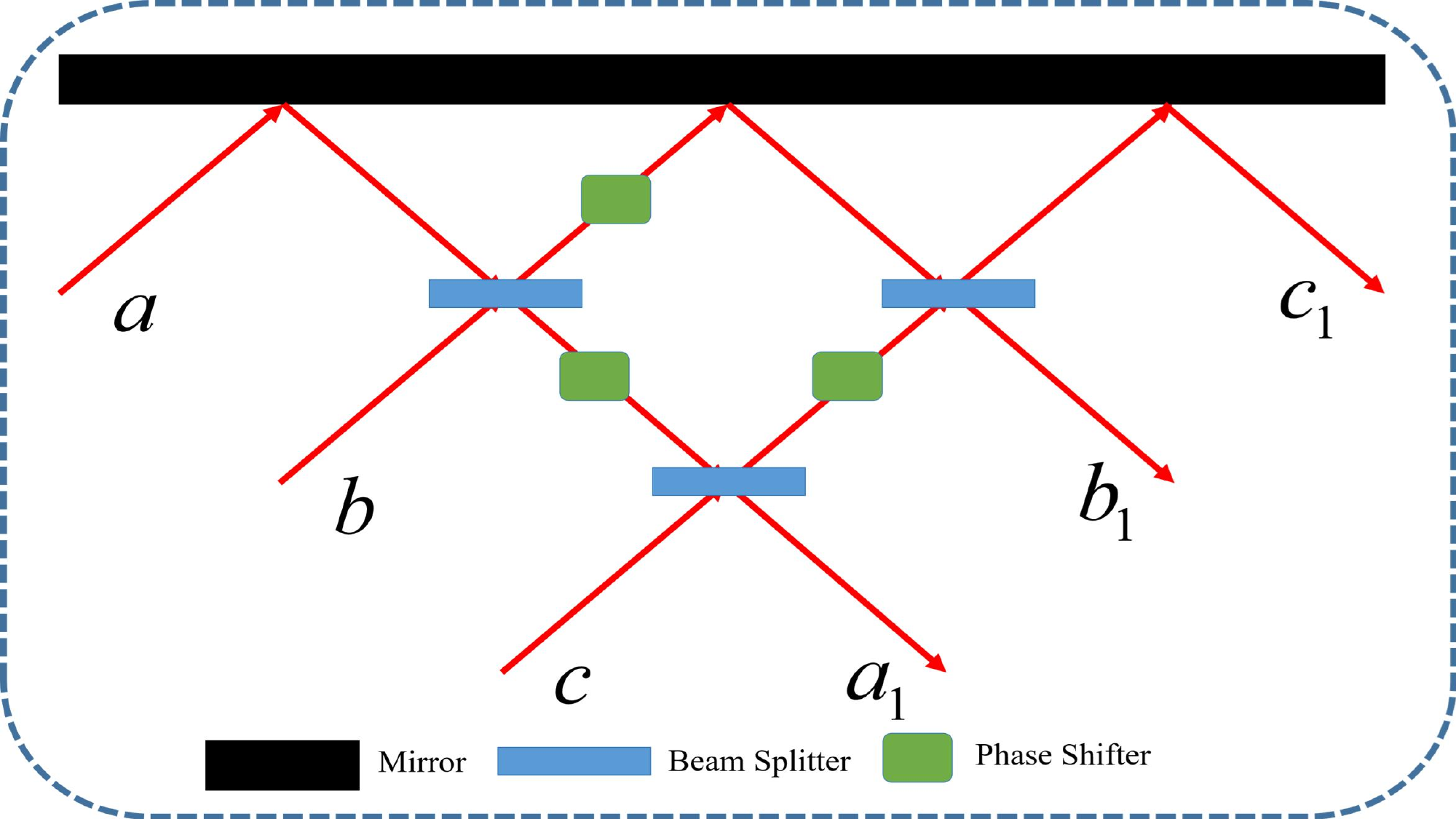}
\caption{ Schematic of the tritter setup. The balanced three-port
interferometer mixes the input modes. Driven by a two-mode squeezed vacuum
in modes $a,b$ and a coherent state in mode $c$, it generates a tripartite
Gaussian output state whose correlation properties are the focus of this
work. }
\label{fig:tritter}
\end{figure}

\subsection{Covariance Matrix Formalism}

For Gaussian states, all information about quantum correlations is contained
in the first and second statistical moments. Since displacements do not
affect entanglement or steering measures for Gaussian states, we focus on
the covariance matrix (CM) $\bm{V}$. For a three-mode system with quadrature
vector $\bm{\xi} = (\hat{x}_a, \hat{p}_a, \hat{x}_b, \hat{p}_b, \hat{x}_c,
\hat{p}_c)^\top$, the CM elements are $V_{ij} = \frac{1}{2} \langle \{\Delta
\hat{\xi}_i, \Delta \hat{\xi}_j\} \rangle$, where $\Delta \hat{\xi}_i = \hat{%
\xi}_i - \langle \hat{\xi}_i \rangle$.

The CM of the output state $|\psi_{\text{out}}\rangle$ is derived
analytically (see Appendix~\ref{app:cov}):
\begin{equation}
\bm{V} = \frac{\bm{I}_6}{2} + \frac{1}{1-\lambda^2}
\begin{pmatrix}
\bm{V}_{11} & \bm{V}_{12} & \bm{V}_{13} \\
\bm{V}_{12}^\top & \bm{V}_{22} & \bm{V}_{23} \\
\bm{V}_{13}^\top & \bm{V}_{23}^\top & \bm{V}_{33}%
\end{pmatrix}%
,  \label{eq:CM_ideal}
\end{equation}
where $\lambda = \tanh r$, and the $2\times2$ submatrices $\bm{V}_{ij}$ are
given by:
\begin{align}
\bm{V}_{11} &= \bm{V}_{22} = \bm{V}_{33} = \frac{\lambda}{3}
\begin{pmatrix}
2\lambda - 1 & -\sqrt{3} \\
-\sqrt{3} & 2\lambda + 1%
\end{pmatrix}%
, \\
\bm{V}_{12} &= \frac{\lambda}{6}
\begin{pmatrix}
1-2\lambda & \sqrt{3} \\
\sqrt{3} & -1-2\lambda%
\end{pmatrix}%
, \\
\bm{V}_{13} &= \bm{V}_{23} = \frac{\lambda}{6}
\begin{pmatrix}
\lambda - 2 & -\sqrt{3}\lambda \\
\sqrt{3}\lambda & \lambda + 2%
\end{pmatrix}%
.
\end{align}
Crucially, the CM is independent of $\gamma$, confirming that displacement
does not alter the correlation structure.

\subsection{Quantifiers for Entanglement and Steering}

For a bipartite split of a Gaussian state with CM $\bm{V}_{AB}$,
entanglement can be detected via the positive partial transpose (PPT)
criterion \cite{Simon2000,Adesso2006}. The logarithmic negativity $E^{A|B}$
quantifies entanglement \cite{Adesso2007}:
\begin{equation}
E^{A|B} = \max\left[ 0, -\sum_{j: \nu_j^{AB^\top} < 1/2}
\ln(2\nu_j^{AB^\top}) \right],  \label{eq:log_neg}
\end{equation}
where $\{\nu_j^{AB^\top}\}$ are the symplectic eigenvalues of the partially
transposed CM.

Gaussian EPR steering from party $A$ to party $B$ is quantified by \cite%
{Kogias2015}:
\begin{equation}
S^{A \rightarrow B} = \max\left[ 0, -\sum_{j: \bar{\nu}_j^{B|A} < 1/2} \ln(2%
\bar{\nu}_j^{B|A}) \right],  \label{eq:steer_quant}
\end{equation}
where $\{\bar{\nu}_j^{B|A}\}$ are the symplectic eigenvalues of the Schur
complement $\bm{V}^{B|A} = \bm{V}_B - \bm{V}_{C}^\top \bm{V}_{A}^{-1} \bm{V}%
_{C}$, with $\bm{V}_{AB}$ partitioned as:
\begin{equation}
\bm{V}_{AB} =
\begin{pmatrix}
\bm{V}_A & \bm{V}_C \\
\bm{V}_C^\top & \bm{V}_B%
\end{pmatrix}%
.
\end{equation}
A nonzero value indicates steerability from $A$ to $B$. The measure is
asymmetric: $S^{A \rightarrow B} \neq S^{B \rightarrow A}$ in general. For
completeness, we will also employ the Gaussian R\'{e}nyi-2 entropy-based
entanglement and steering monotones in Sec.~\ref{sec:discussion}, which
offer a complementary perspective and confirm the robustness of our main
findings.

\section{Quantum Correlations under Ideal Conditions}

\label{sec:ideal_correlations}

\subsection{Entanglement Structure}

\begin{figure}[tbh]
\centering
\includegraphics[width=0.8\columnwidth]{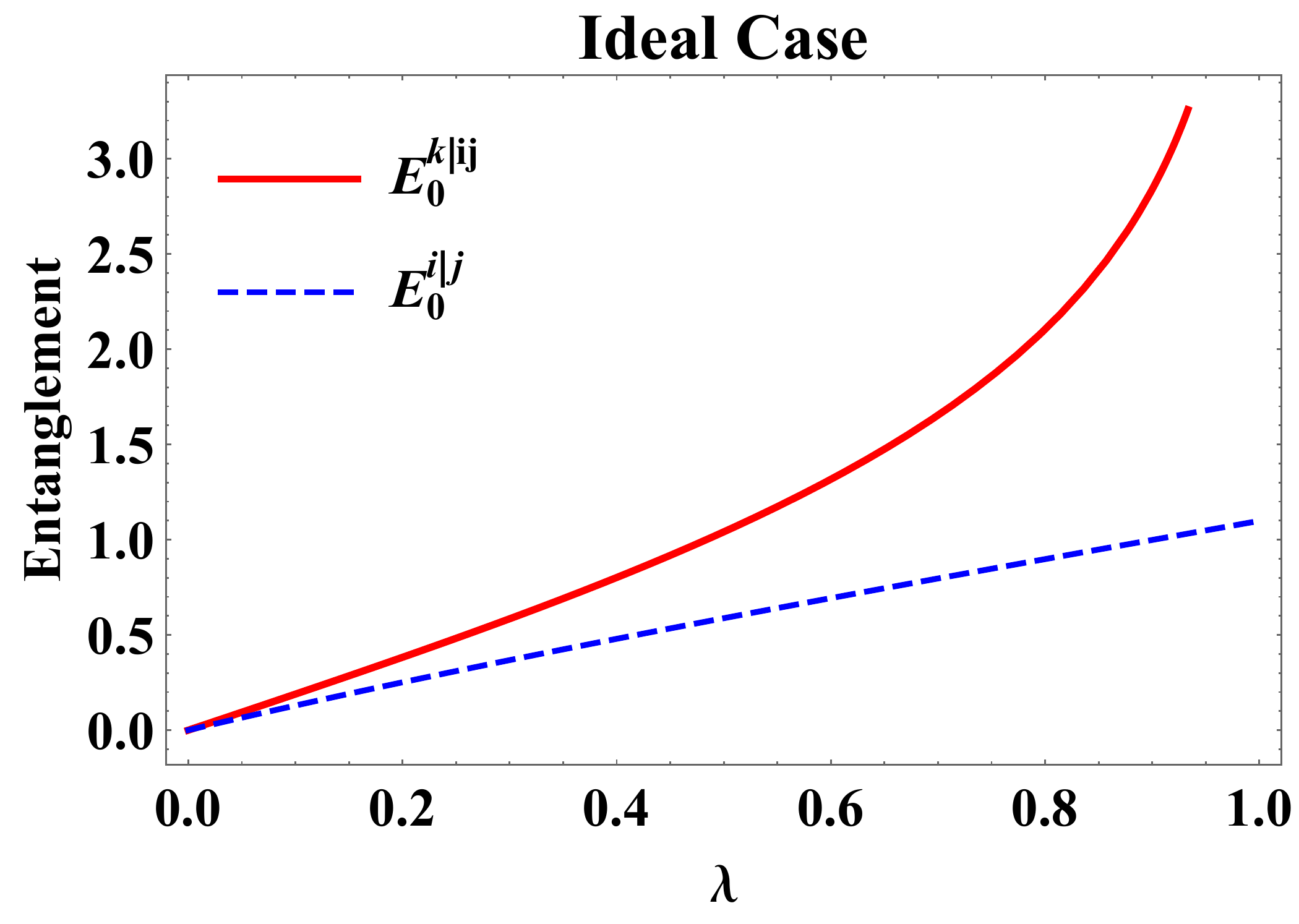}
\caption{Entanglement versus squeezing parameter $\protect\lambda$ in the
lossless case. The tripartite entanglement $E^{k|ij}$ (blue dashed)
consistently dominates the bipartite entanglement $E^{i|j}$ (red dotted) for
any $\protect\lambda>0$. The monotonic increase reflects the growing quantum
resource. }
\label{fig:E_vs_lambda}
\end{figure}

Using Eq.~(\ref{eq:log_neg}), we compute the bipartite entanglement for all
possible splits. For any two modes $i$ and $j$ ($i,j\in \{a,b,c\}$), the
bipartite entanglement is:
\begin{equation}
E^{i|j}=\ln \left[ \frac{3(1+\lambda )}{3-\lambda }\right] .
\label{eq:E_2mode}
\end{equation}
Remarkably, this value is identical for any pair $(i,j)$ because the ideal
tritter symmetrizes the input TMSV correlations across all three output
modes.For the split between one mode $k$ and the remaining two modes $ij$,
the tripartite entanglement is:
\begin{equation}
E^{k|ij}=\ln \left[ \frac{9(1-\lambda ^{2})}{(\sqrt{9-\lambda ^{2}}-\sqrt{8}%
\lambda )^{2}}\right] .  \label{eq:E_1vs2}
\end{equation}
Both measures are positive for $\lambda >0$ and increase monotonically with $%
\lambda $, as shown in Fig.~\ref{fig:E_vs_lambda}. Moreover, $%
E^{k|ij}>E^{i|j}$ for all $\lambda >0$, indicating stronger entanglement in
the $1$ vs $2$ partition. The nonzero entanglement across all bipartitions
confirms that the output state is a genuinely tripartite entangled Gaussian
state. Because the input TMSV lives in modes $a,b$, one might expect an
asymmetry between $c$ and the other modes. However, the symmetric tritter
evenly distributes the correlations, making all three $1$ vs $2$
entanglements equal.

\subsection{EPR Steering Hierarchy}

Applying Eq.~(\ref{eq:steer_quant}), we find a starkly different structure
for steering. The steering measures between any two individual modes vanish
identically:
\begin{equation}
S^{i \leftrightarrow j} = 0, \quad \forall i,j.  \label{eq:S_2mode_zero}
\end{equation}
This zero pairwise steering is a direct consequence of the tritter's
symmetric mixing: no single mode retains enough asymmetric information to
steer another individual mode. However, steering exists between a single
mode and the group of the other two:
\begin{equation}
S^{k \rightarrow ij} = S^{ij \rightarrow k} = \ln\left[ \frac{9-\lambda^2}{%
9(1-\lambda^2)} \right] > 0 \quad \text{for } \lambda > 0.  \label{eq:S_1vs2}
\end{equation}
This demonstrates a strict hierarchy: while all modes are pairwise
entangled, no pairwise steering exists. Steering is only present in the
collective $1$ vs $2$ partition, and it is \emph{symmetric} ($S^{k
\rightarrow ij} = S^{ij \rightarrow k}$) in the ideal case.

The monogamy of Gaussian steering \cite{Yu2017, Zhang2024} imposes
constraints:
\begin{equation}
S^{ij \rightarrow k} - S^{i \rightarrow k} - S^{j \rightarrow k} \geq 0,
\quad S^{k \rightarrow ij} - S^{k \rightarrow i} - S^{k \rightarrow j} \geq
0.  \label{eq:monogamy}
\end{equation}
Given $S^{i \rightarrow k} = S^{k \rightarrow i} = 0$, these inequalities
are trivially satisfied, indicating no sharing of steering exists in the
ideal output state.

\subsection{Physical Mechanism and Interpretation}

Ideal conditions reveal the phenomenon of ``entanglement without steering''
between any two modes, which has profound physical origins. The symmetric
tritter operation distributes the quadrature squeezing uniformly among all
three modes, thereby washing out any directional bias. As a result, although
non-classical correlations survive, no single party has sufficient
informational leverage to steer another individual party. This is a clear
signature of the resource competition in multipartite systems.

From a phase-space perspective, the tritter's unitary transformation (Eq.~%
\ref{eq:tritter_U}) mixes the quadratures of all three modes symmetrically.
This symmetric mixing dilutes any directional dependence in the
correlations, making it impossible for one mode to unilaterally determine
the state of another. The preservation of entanglement, however, indicates
that non-classical correlations remain, albeit in a symmetric form that
cannot be exploited for steering.

From a resource theory perspective \cite{Li2023}, steering requires not only
non-classical correlations but also sufficient control of one party over the
state of another. In the three-mode system, despite entanglement between any
two modes, the involvement of the third mode prevents any single party from
completely determining the state of another---a direct manifestation of
correlation sharing and resource competition in multipartite systems. Our
calculations align with the monogamy constraints of multipartite steering
\cite{Yu2017, Zhang2024} and provide quantitative tools for understanding
correlation distribution in multipartite systems.

\section{Robustness against Loss}

\label{sec:loss}

\subsection{Loss Model}

\begin{figure}[tbh]
\centering
\includegraphics[width=0.8\columnwidth]{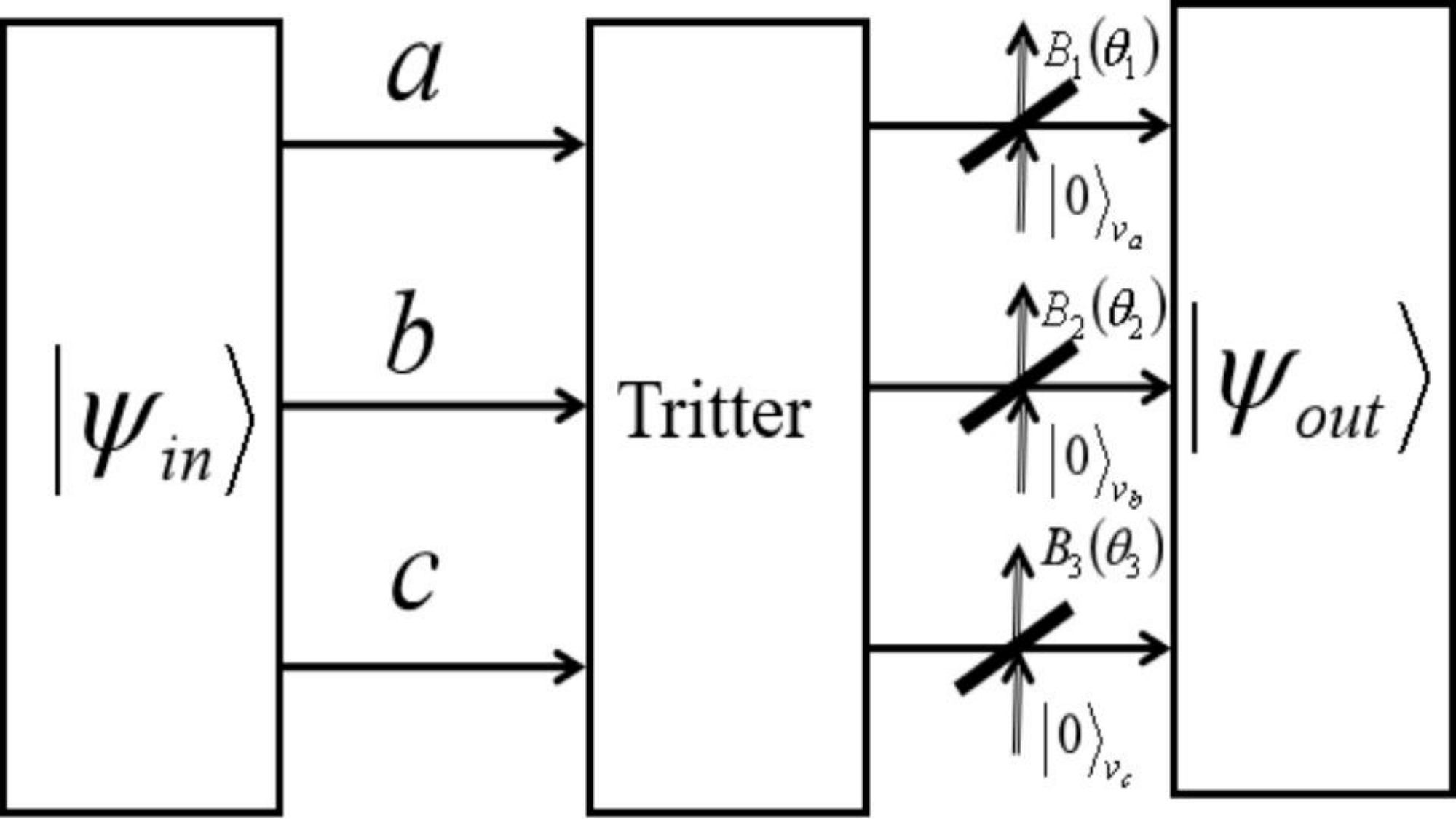}
\caption{ Loss model. After the tritter, each output mode undergoes an
independent loss channel of transmissivity $T_i$. The covariance matrix is
transformed accordingly. }
\label{fig:loss_model}
\end{figure}

To model realistic conditions, we incorporate loss channels in each output
mode of the tritter, as depicted in Fig.~\ref{fig:loss_model}. Each channel
is characterized by a transmissivity $T_{i}=\cos ^{2}\theta _{i}$ ($i=1,2,3$%
), with $T_{i}=1$ representing no loss. The effect on the CM is a simple
scaling: the CM $\bm{V}^{\text{lossy}}$ is obtained from the ideal CM $\bm{V}
$ [Eq.~(\ref{eq:CM_ideal})] by the transformation $\bm{V}_{ij}\rightarrow
\sqrt{T_{i}T_{j}}\,\bm{V}_{ij}$ for $i\neq j$, and $\bm{V}_{ii}\rightarrow
T_{i}\bm{V}_{ii}+(1-T_{i})\bm{I}_{2}/2$ \cite{Zheng2026}.

We analyze five representative loss configurations (Scenarios 1--5) to
understand the effect of asymmetric versus symmetric loss on the correlation
hierarchy \cite{Liu2023, Kumar2022, Lami2018}. These scenarios are defined
as follows:

\begin{itemize}
\item \textbf{Scenario 1: Single-mode loss} - Only mode $k$ experiences
loss, $T_i = T_j = 1$, $T_k = T$.

\item \textbf{Scenario 2: Loss at one mode of the pair} - Only one mode in
the dual-mode module experiences loss, $T_i = T$, $T_j = T_k = 1$ or $T_j =
T $, $T_i = T_k = 1$.

\item \textbf{Scenario 3: Group-matched dual-mode loss} - Both modes in the
dual-mode module experience loss, $T_i = T_j = T$, $T_k = 1$.

\item \textbf{Scenario 4: Intergroup dual-mode loss} - One mode in the pair
and mode $k$ experience loss, $T_i = T_k = T$, $T_j = 1$ or $T_j = T_k = T$,
$T_i = 1$.

\item \textbf{Scenario 5: Triple-mode loss} - All three modes experience
loss, $T_i = T_j = T_k = T$.
\end{itemize}

These configurations cover the physically distinct ways in which loss can
couple to the modes, thereby providing a complete picture of how the spatial
distribution of noise affects the various quantum correlations.

\subsection{Evolution of Quantum Correlations under Loss}

\subsubsection{Entanglement}

Based on the entanglement metric defined in Eq.~(\ref{eq:log_neg}), the
bipartite entanglement under loss is given by:
\begin{align}
E_{1}^{i|j}& =\ln \left[ \frac{9(1-\lambda ^{2})}{9-\delta _{1}\lambda
^{2}-4\lambda \sqrt{\delta _{2}\lambda ^{2}+9T}}\right] ,
\label{eq:E2_loss1} \\
E_{2}^{i|j}& =\ln \left[ \frac{3(1-\lambda ^{2})}{(3-3\lambda +2T\lambda
)(1+\lambda -2T\lambda )}\right] ,  \label{eq:E2_loss2}
\end{align}%
where $\delta _{1}=5-16T+8T^{2}$, $\delta _{2}=(1-T)^{2}(1-8T+4T^{2})$.
Subscript 1 indicates loss in one of the two modes, while subscript 2
indicates loss in both modes.

For tripartite entanglement $E^{k|ij}$, analytical expressions for Scenarios
1, 3, and 5 are:
\begin{align}
E_{1}^{k|ij} &= \ln\left[ \frac{9(1-\lambda^2)}{9 - \epsilon_1\lambda^2 -
4\lambda\sqrt{18T + \epsilon_2\lambda^2}} \right],  \label{eq:E3_loss1} \\
E_{3}^{k|ij} &= \ln\left[ \frac{9(1-\lambda^2)}{9 - \epsilon_3\lambda^2 -
2\lambda\sqrt{72T + \epsilon_4\lambda^2}} \right],  \label{eq:E3_loss3} \\
E_{5}^{k|ij} &= \ln\left[ \frac{9(1-\lambda^2)}{9 - \epsilon_5\lambda^2 -
2T\lambda\sqrt{72 + \epsilon_6\lambda^2}} \right],  \label{eq:E3_loss5}
\end{align}
where $\epsilon_1 = 5-20T+8T^2$, $\epsilon_2 = (1-4T+T^2)(1-2T)^2$, $%
\epsilon_3 = 5-14T+2T^2$, $\epsilon_4 = (1-10T+T^2)(2-T)^2$, $\epsilon_5 =
9-18T+2T^2$, $\epsilon_6 = 9-18T+T^2$.

It is to be noted that the tripartite entanglement expressions for Scenarios
2 and 4,ie $E_{2}^{k|ij}$ and $E_{4}^{k|ij}$ are too complex to provide
exact analytical expressions.

When loss breaks the symmetry, the entanglement for different choices of the
single mode $k$ can differ. For example, in Scenario 4 with $T_a=T_c=T$ and $%
T_b=1$, we find $E^{a|bc} \neq E^{b|ac} \neq E^{c|ab}$. The expressions
above correspond to the specific partition where $k$ is the mode indicated
in each scenario's definition. A complete table of all partition-dependent
expressions is provided in the Supplemental Material.

Both bipartite $E^{i|j}$ and tripartite $E^{k|ij}$ entanglement decrease
monotonically with increasing loss (decreasing $T$), as shown in Figs.~\ref%
{fig:E_loss} and~\ref{fig:E_loss2}. For bipartite entanglement, it is more
fragile when loss affects both modes involved in the partition. Furthermore,
the rate of entanglement decay under loss increases with the squeezing
parameter $\lambda$. Bipartite entanglement abruptly vanishes under
significant loss when $\lambda$ exceeds a certain threshold. Physically,
larger $\lambda$ corresponds to a higher average photon number, which
enhances the interaction with the lossy environment and accelerates
decoherence.

To clarify the impact of loss distribution on the tripartite entanglement of
the output state, we compare the entanglement evolution curves as a function
of reflectivity in the five scenarios, as shown in Fig.~\ref%
{fig:E_comparison}. The figure reveals significant differences in
entanglement strength across the various loss distributions, ranked from
strongest to weakest as follows: $%
E_{2}^{k|ij}>E_{3}^{k|ij}>E_{1}^{k|ij}>E_{4}^{k|ij}>E_{5}^{k|ij}$. This
ordering has a clear physical origin: loss on a single mode of the pair
(Scenario 2) minimally disturbs the dominant bipartite correlations, whereas
symmetric loss on all modes (Scenario 5) is the most destructive.

\begin{figure}[tbh]
\centering
\includegraphics[width=0.8\columnwidth]{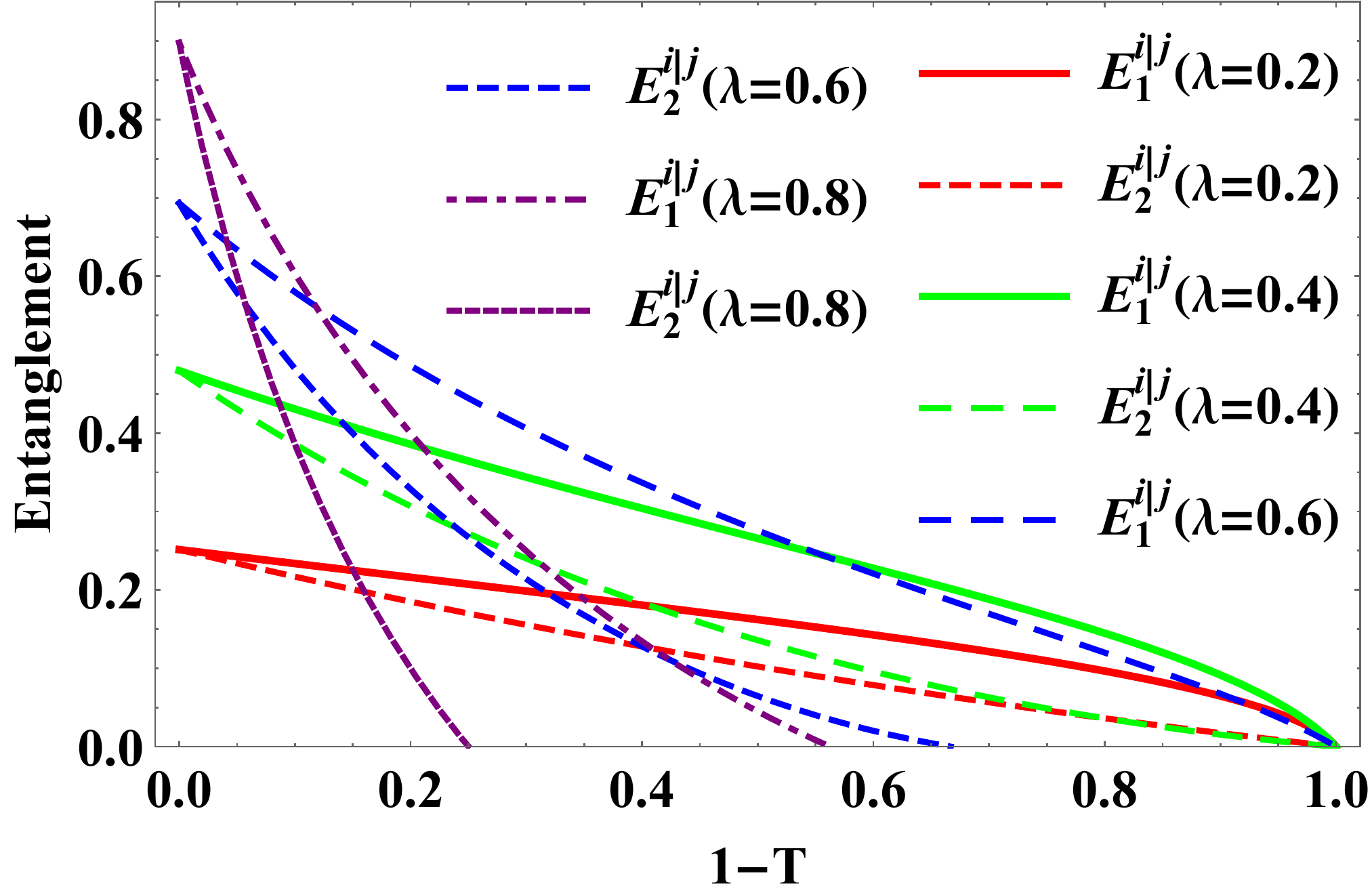}
\caption{ Bipartite entanglement $E^{i|j}$ as a function of reflectivity $%
1-T $ for different $\protect\lambda$. Larger squeezing leads to a sharper
drop and earlier sudden death of entanglement, because higher photon-number
states couple more strongly to the loss reservoir. }
\label{fig:E_loss}
\end{figure}

\begin{figure}[tbh]
\centering
\subfigure[]{\includegraphics[width=0.83\columnwidth]{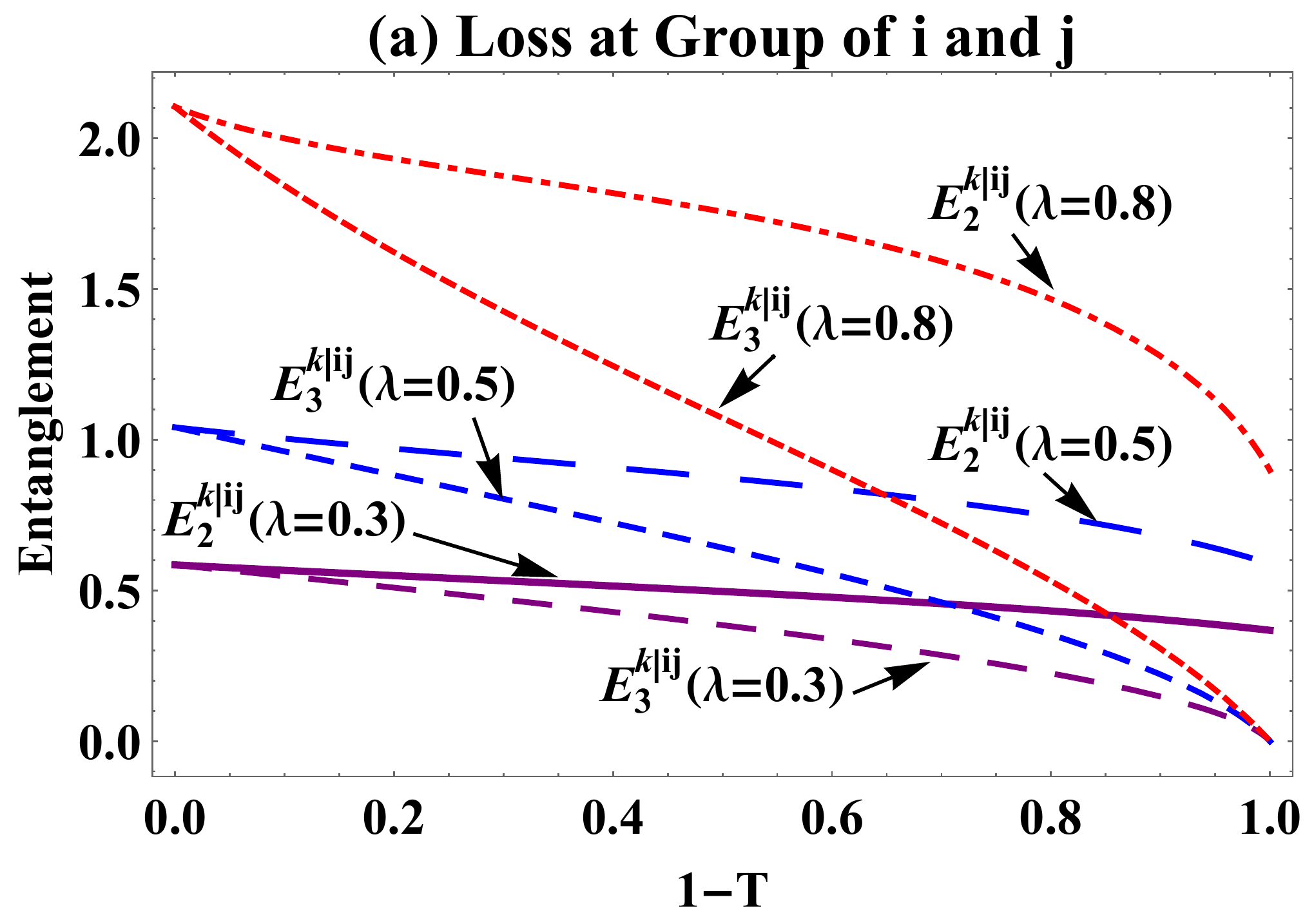}} \newline
\subfigure[]{\includegraphics[width=0.83\columnwidth]{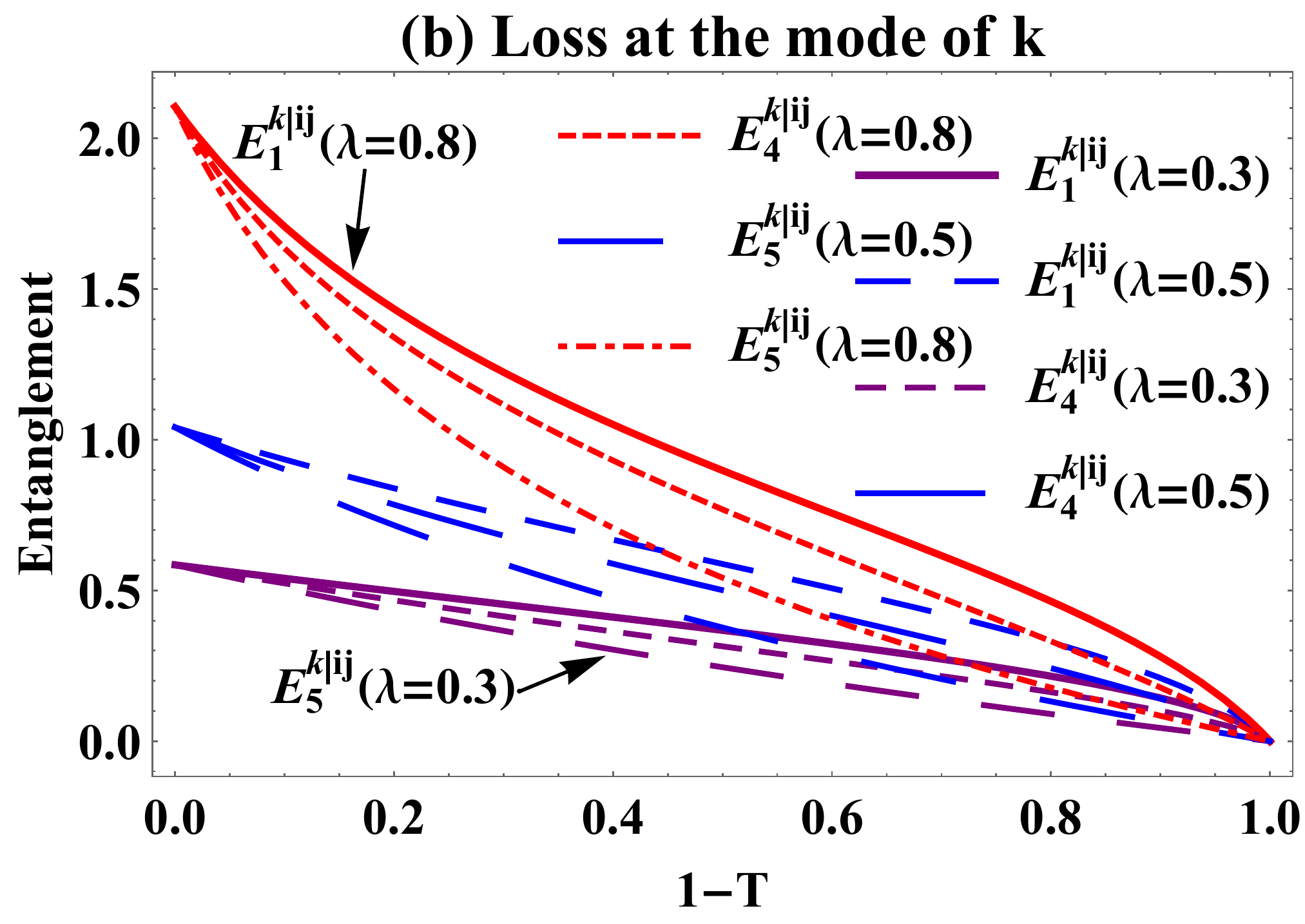}}
\caption{ Tripartite entanglement $E^{k|ij}$ versus reflectivity for
different loss configurations. (a) Scenarios with no loss in mode $k$; (b)
scenarios with loss in mode $k$. The qualitative behavior highlights the
role of the loss location: when mode $k$ is lossless, entanglement decays
more gently. }
\label{fig:E_loss2}
\end{figure}

\begin{figure}[tbh]
\centering
\subfigure[]{\includegraphics[width=0.83\columnwidth]{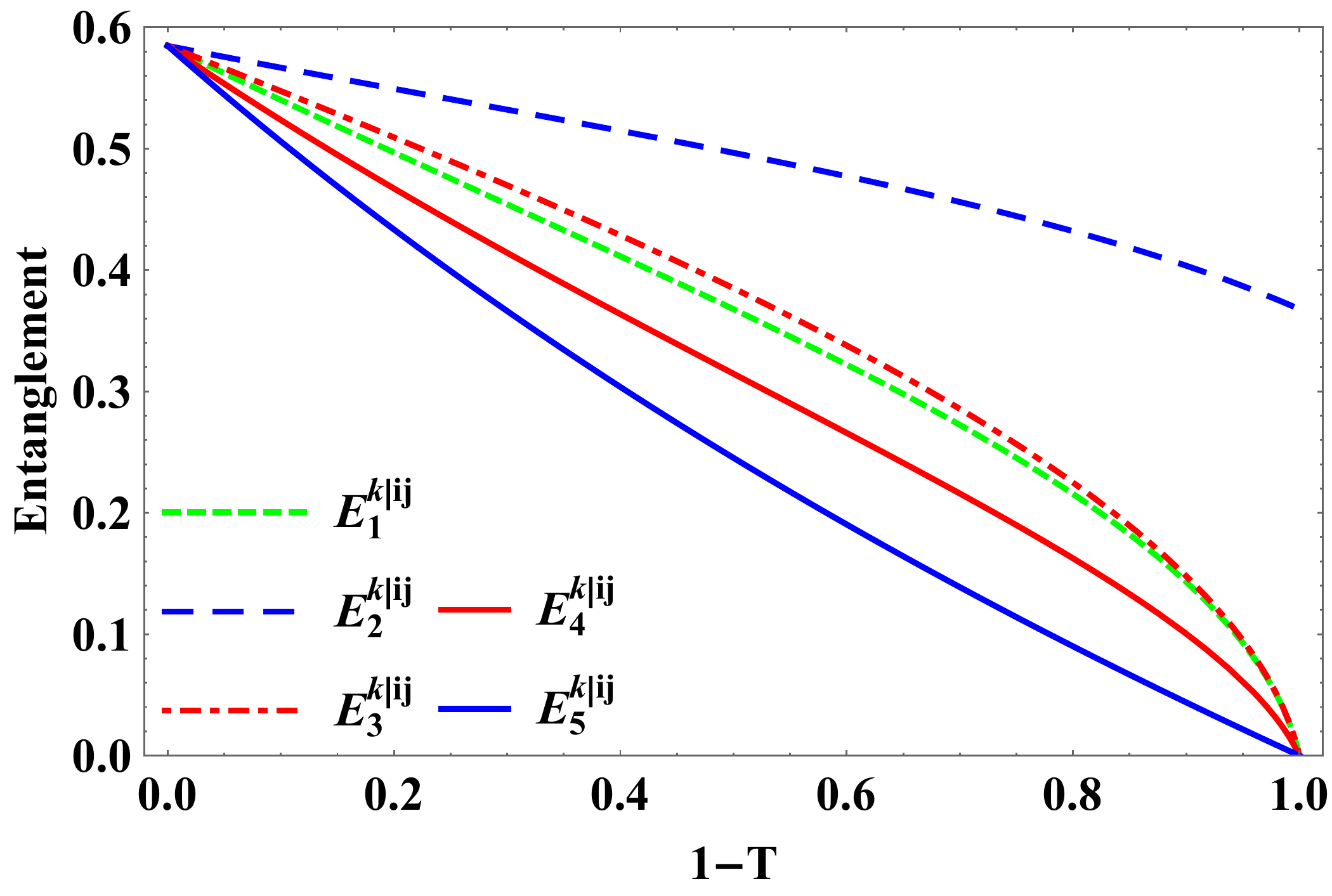}} \newline
\subfigure[]{\includegraphics[width=0.83\columnwidth]{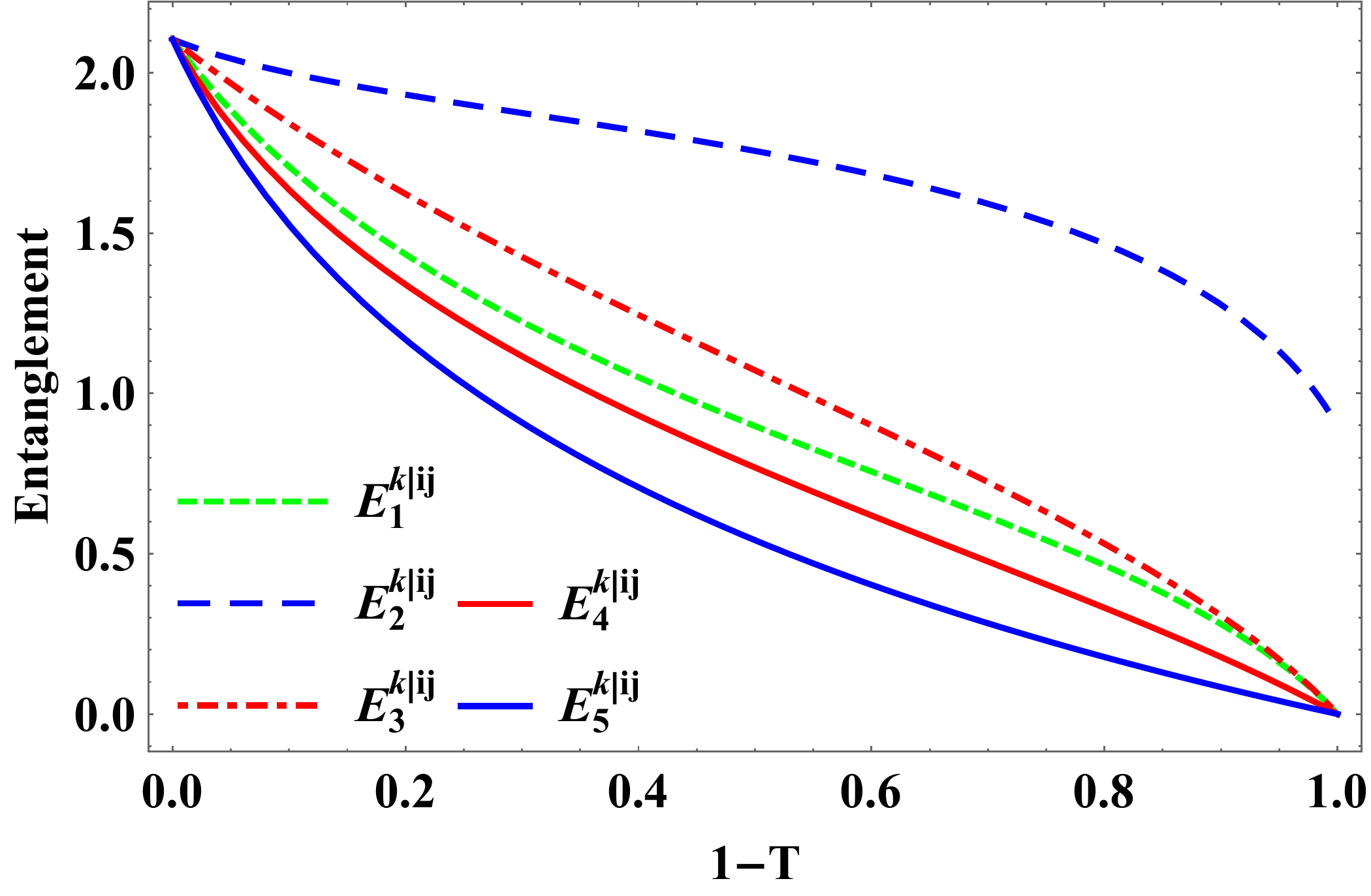}}
\caption{ Comparison of tripartite entanglement $E^{k|ij}$ across all five
loss scenarios for (a) $\protect\lambda=0.3$ and (b) $\protect\lambda=0.8$.
The hierarchy $E_2>E_3>E_1>E_4>E_5$ is clearly visible and is explained by
the degree to which loss affects the correlated modes. }
\label{fig:E_comparison}
\end{figure}

\subsubsection{Steering}

The steerability $S^{i\rightarrow j}$ between any two modes $i$ and $j$ can
be calculated using the symplectic eigenvalues $v^{j|i}$ of the Schur
complement $V^{j|i}$. In the loss model, the corresponding symplectic
eigenvalues can be expressed as:
\begin{equation}
v_{1}^{j|i}=v_{2}^{j|i}=\sqrt{\frac{9-2\xi _{1}\lambda ^{2}+\xi
_{2}^{2}\lambda ^{4}}{4(1-\lambda ^{2})\left[ 9-(3-4T_{i})^{2}\lambda ^{2}%
\right] }},  \label{eq:sv_loss}
\end{equation}%
where $\xi _{1}=9+8(T_{i}^{2}+T_{j}^{2})-12(T_{i}+T_{j})+4T_{i}T_{j}$, $\xi
_{2}=3-4(T_{i}+T_{j}-T_{i}T_{j})$. Since $v_{1}^{j|i}=v_{2}^{j|i}\geq \frac{1%
}{2}$, according to the definition of the steering metric in Eq.~(\ref%
{eq:steer_quant}), no quantum steering relationship exists between any two
modes. Thus, the system satisfies the steering monogamy relation under loss.
This finding, that pairwise steerability remains exactly zero even under
asymmetric loss, is non-trivial and underscores the robustness of the
tritter's symmetrization against local noise with respect to bipartite
steering.

Similarly, tripartite steering in the five scenarios is given by:

\textbf{Scenario 1}:
\begin{align}
S_{1}^{ij\rightarrow k} &= \ln\left[ \frac{9-\lambda^2}{9-(1-4T)^2\lambda^2} %
\right], \\
S_{1}^{k\rightarrow ij} &= \ln\left[ \frac{9-(3-4T)^2\lambda^2}{%
9-(1-4T)^2\lambda^2} \right].
\end{align}

\textbf{Scenario 2}:
\begin{align}
S_{2}^{ij\rightarrow k} &= \ln\left[ \frac{9-2(5-8T+8T^2)\lambda^2+\lambda^4%
}{(1-\lambda^2)[9-(1-4T)^2\lambda^2]} \right], \\
S_{2}^{k\rightarrow ij} &= \ln\left[ \frac{(1-\lambda^2)(9-\lambda^2)}{%
(1-\lambda^2)(9-\lambda^2) + \chi_1 - 4\lambda^2\sqrt{T\chi_2}} \right],
\end{align}
where $\chi_1 = 4T\lambda^2(1+\lambda^2-2T)$, $\chi_2 =
(1-\lambda^2)(9-\lambda^2) + 4T^3 - 4T^2(1+\lambda^2) - 4T(2-3\lambda^2)$.

\textbf{Scenario 3}:
\begin{align}
S_{3}^{ij\rightarrow k} &= \ln\left[ \frac{9-(3-2T)^2\lambda^2}{%
9-(1+2T)^2\lambda^2} \right], \\
S_{3}^{k\rightarrow ij} &= \ln\left[ \frac{9-\lambda^2}{9-(1+2T)^2\lambda^2} %
\right].
\end{align}

\textbf{Scenario 4}:
\begin{align}
S_{4}^{ij\rightarrow k}& =\ln \left[ \frac{9-2(5-8T+8T^{2})\lambda
^{2}+\lambda ^{4}}{9-2(5-16T+20T^{2})\lambda ^{2}+(1-4T^{2})^{2}\lambda ^{4}}%
\right] , \\
S_{4}^{k\rightarrow ij}& =\ln \left[ \frac{(1-\lambda
^{2})[9-(3-4T)^{2}\lambda ^{2}]}{9+\vartheta _{0}\lambda ^{4}-2\lambda ^{2}%
\left[ \vartheta _{1}+2\sqrt{\vartheta _{2}+\vartheta _{3}+\vartheta
_{4}+\vartheta _{5}}\right] }\right] ,
\end{align}%
with $\vartheta _{0}=5-20T+28T^{2}-16T^{3}+8T^{4}$, $\vartheta
_{1}=7-14T+14T^{2}$, $\vartheta _{2}=4T^{8}\lambda ^{4}-16T^{7}\lambda ^{4}$%
, $\vartheta _{3}=(1-\lambda ^{2})^{2}-4T^{6}\lambda ^{2}(3-7\lambda
^{2})+T^{5}\lambda ^{2}(60-52\lambda ^{2})$, $\vartheta _{4}=T(-7+18\lambda
^{2}-11\lambda ^{4})-4T^{3}(6-29\lambda ^{2}+23\lambda ^{4})$, $\vartheta
_{5}=T^{2}(22-64\lambda ^{2}+46\lambda ^{4})+T^{4}(9-118\lambda
^{2}+93\lambda ^{4})$.

\textbf{Scenario 5}:
\begin{align}
S_{5}^{ij\rightarrow k} &= \ln\left[ \frac{9-(3-2T)^2\lambda^2}{%
9-(3-6T)^2\lambda^2} \right], \\
S_{5}^{k\rightarrow ij} &= \ln\left[ \frac{9-(3-4T)^2\lambda^2}{%
9-(3-6T)^2\lambda^2} \right].
\end{align}

According to Eqs.~(26)--(35), we can derive the conditions for steering in
the five scenarios:
\begin{align}
S_{1}^{ij\rightarrow k}>0& :T>0.5,\quad S_{1}^{k\rightarrow ij}>0:T>0.5, \\
S_{2}^{ij\rightarrow k}>0& :0<T\leq 1,\quad S_{2}^{k\rightarrow
ij}>0:0<T\leq 1, \\
S_{3}^{ij\rightarrow k}>0& :T>0.5,\quad S_{3}^{k\rightarrow ij}>0:0<T\leq 1,
\\
S_{4}^{ij\rightarrow k}>0& :T>2/3,\quad S_{4}^{k\rightarrow ij}>0:T>0.5, \\
S_{5}^{ij\rightarrow k}>0& :T>3/4,\quad S_{5}^{k\rightarrow ij}>0:T>3/5.
\end{align}

These threshold expressions constitute a central result of this work: they
provide a direct operational guide for the transmission requirements needed
to maintain steerability in a given network configuration.

To explicitly demonstrate the impact of asymmetry on different partitions,
consider Scenario 4 with $T_a=T_c=T$, $T_b=1$. The steering from mode $a$ to
the pair $bc$ survives for $T>0.5$, while steering from mode $b$ to the pair
$ac$ requires $T>2/3$. This difference arises because mode $b$, being
lossless, retains higher purity and can thus steer the lossy pair more
effectively than a lossy mode can steer a mixed collective. Such
partition-dependent asymmetries are fully characterized for all scenarios in
the Supplemental Material.

\begin{figure}[tbh]
\centering
\subfigure[]{\includegraphics[width=0.83\columnwidth]{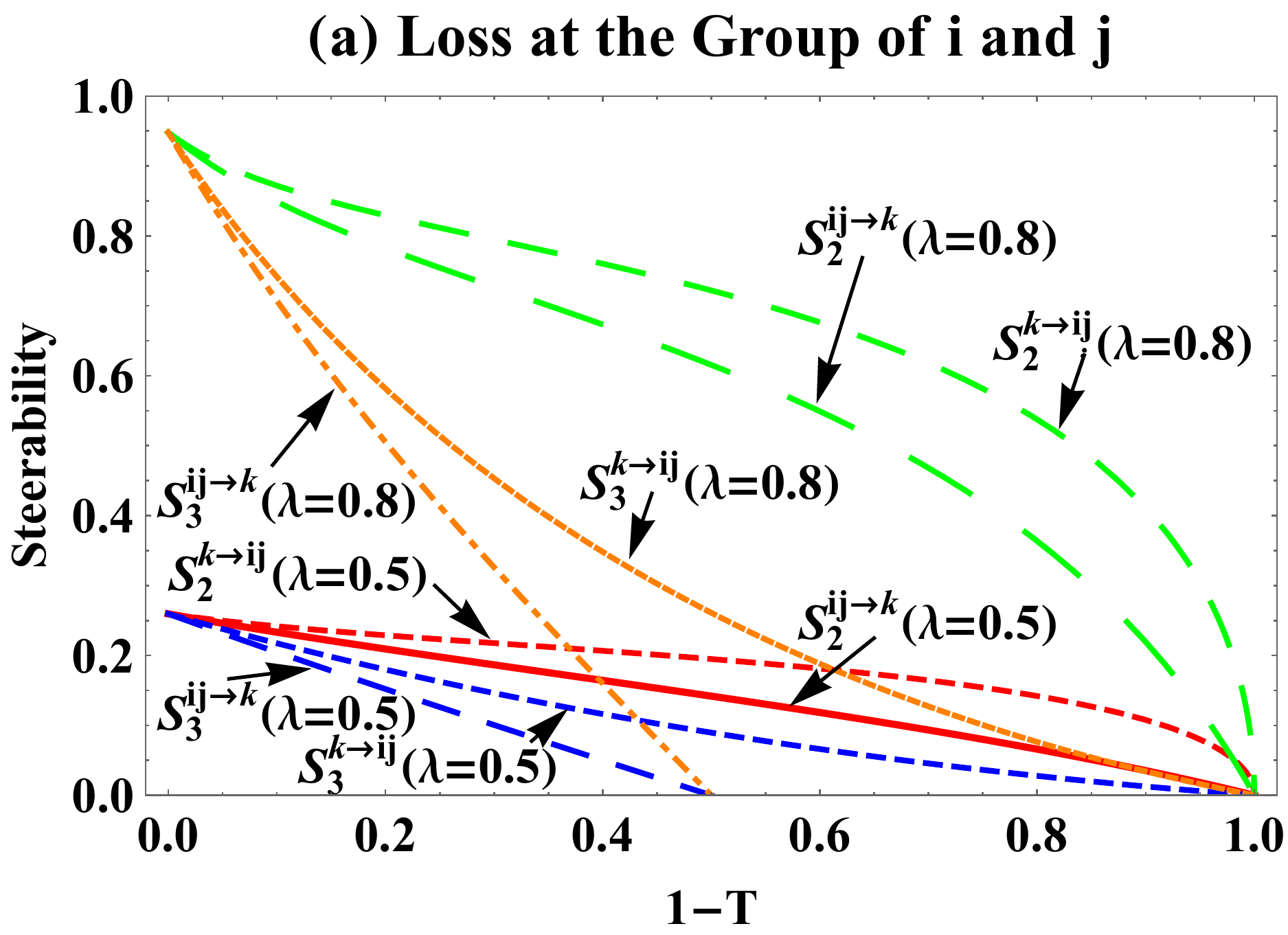}} \newline
\subfigure[]{\includegraphics[width=0.83\columnwidth]{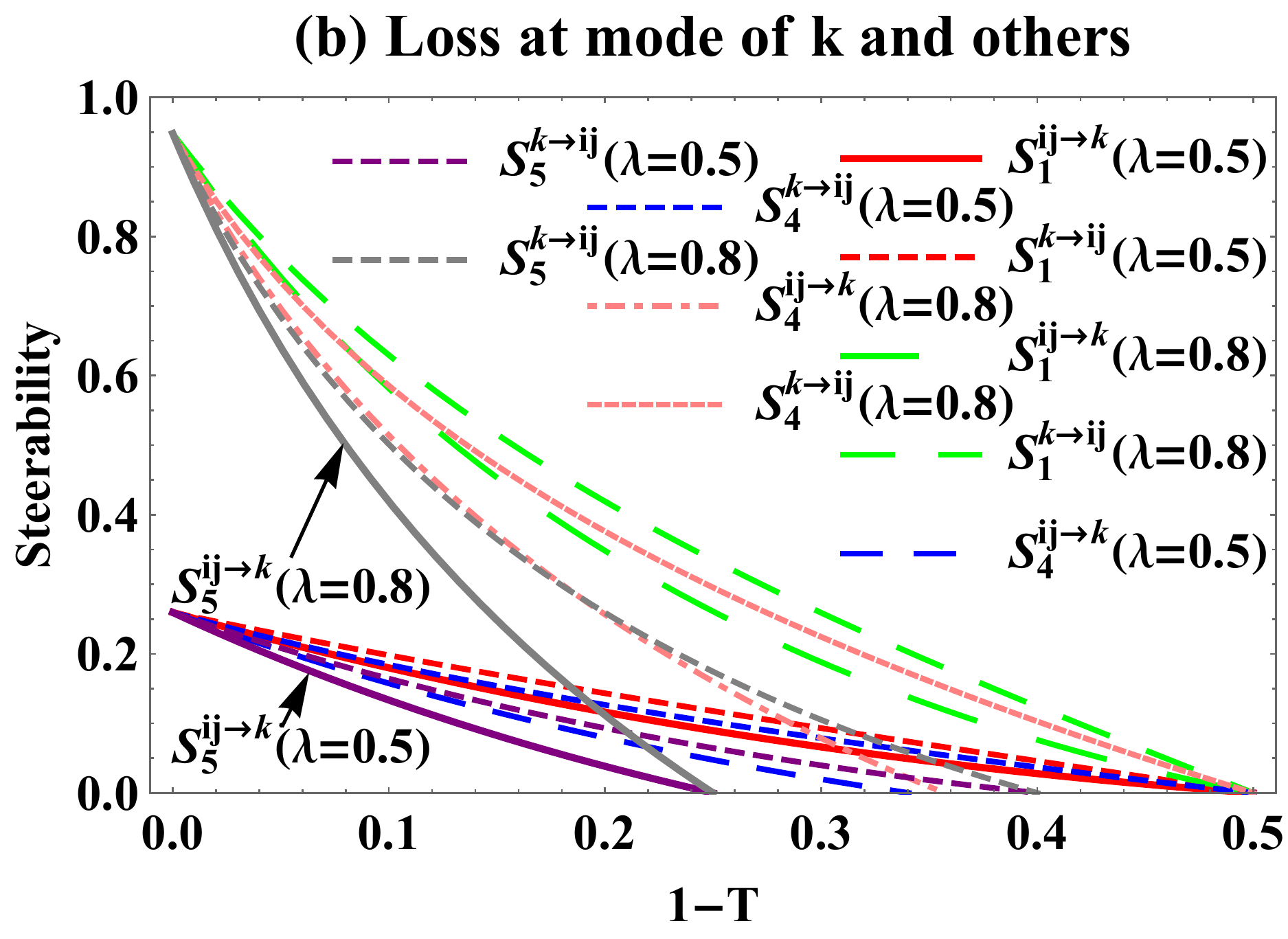}}
\caption{ Steering as a function of reflectivity under different loss
configurations. (a) Scenarios where the steering party is lossless; (b)
scenarios where the steering party experiences loss. The direction $S^{k\to
ij}$ is generally more robust, highlighting the operational advantage of
using a single mode as the steerer. }
\label{fig:S_loss}
\end{figure}

Tripartite steering exhibits a more nuanced and stricter behavior. The
symmetric steering $S^{k\rightarrow ij} = S^{ij\rightarrow k}$ present in
the ideal case is broken by loss, leading to directional asymmetry. More
critically, steering vanishes at finite transmissivity thresholds, which
depend on the loss configuration. Fig.~\ref{fig:S_loss} plots the steering
measures versus loss for different scenarios. Key observations:

\begin{itemize}
\item Directional asymmetry: Steering from a single mode to the other two
modes is generally more robust than the reverse direction, because the
steered party (the collective mode) can better tolerate losses on its
components.

\item Threshold hierarchy: The thresholds for $S^{ij\rightarrow k} > 0$ are
typically stricter than for $S^{k\rightarrow ij} > 0$, except in highly
symmetric configurations.

\item Scenario 2 robustness: In Scenario 2 (loss only in one of modes $i$
and $j$), steering persists for all $T > 0$, demonstrating the highest
robustness. This suggests a practical ``partial protection'' strategy: by
keeping one mode of the pair lossless, steering in both directions can be
maintained even under severe noise on the other mode.

\item Symmetric loss vulnerability: In Scenario 5 (symmetric loss in all
modes), steering vanishes at the highest thresholds, making it the most
fragile configuration.
\end{itemize}

To investigate the influence of loss distributions on steering, we plot the
steering curve as a function of reflectivity $1-T$ in Fig.~\ref%
{fig:S_comparison}. The results reveal significant differences in steering
intensity across the five scenarios, ranked from strongest to weakest as
follows: Scenario 2, Scenario 3, Scenario 1, Scenario 4, and Scenario 5.
This ranking mirrors that of entanglement, confirming that the spatial
distribution of loss is a more decisive factor than the mere total amount of
loss.

\begin{figure}[tbh]
\centering
\subfigure[]{\includegraphics[width=0.83\columnwidth]{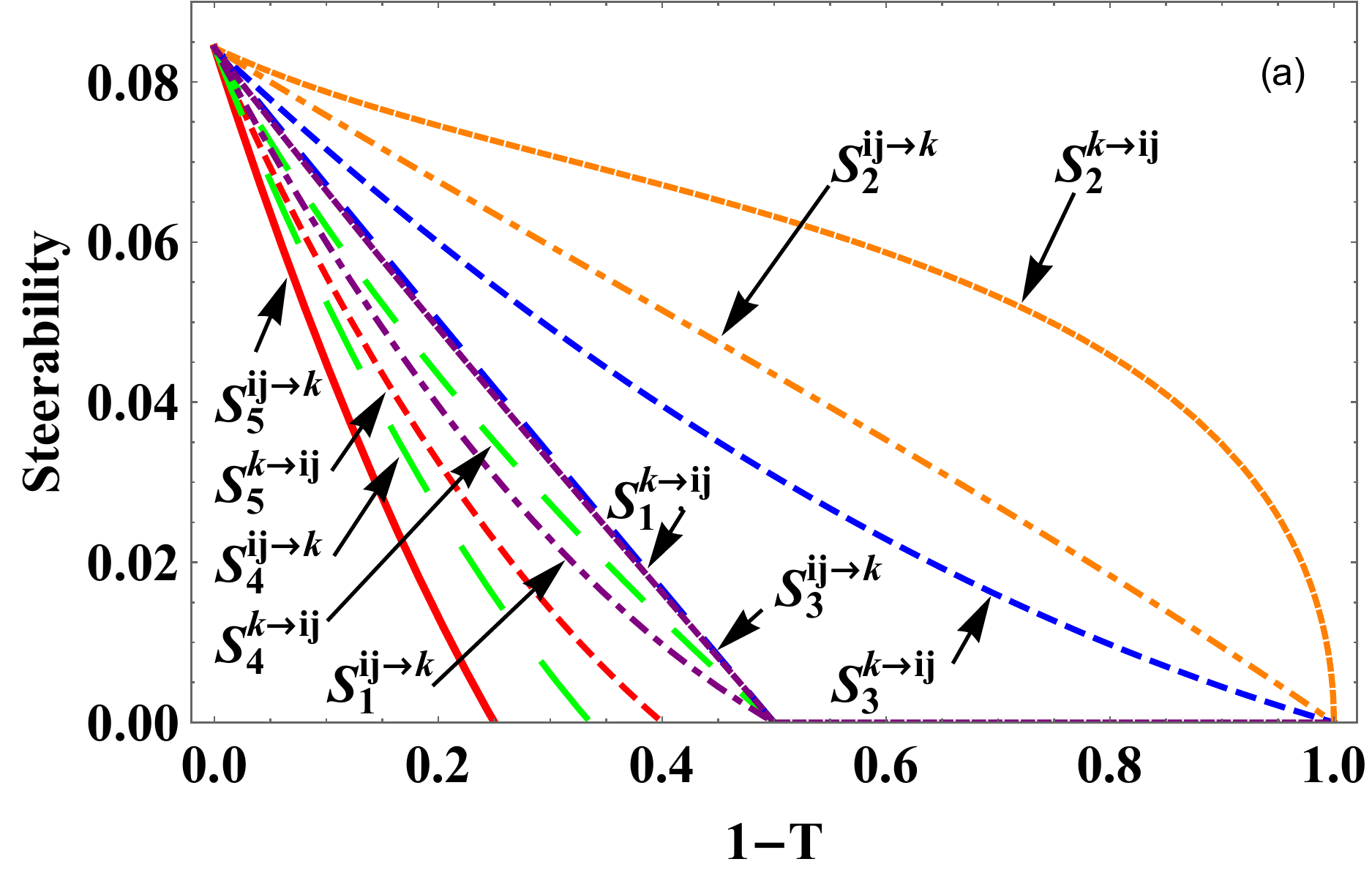}} \newline
\subfigure[]{\includegraphics[width=0.83\columnwidth]{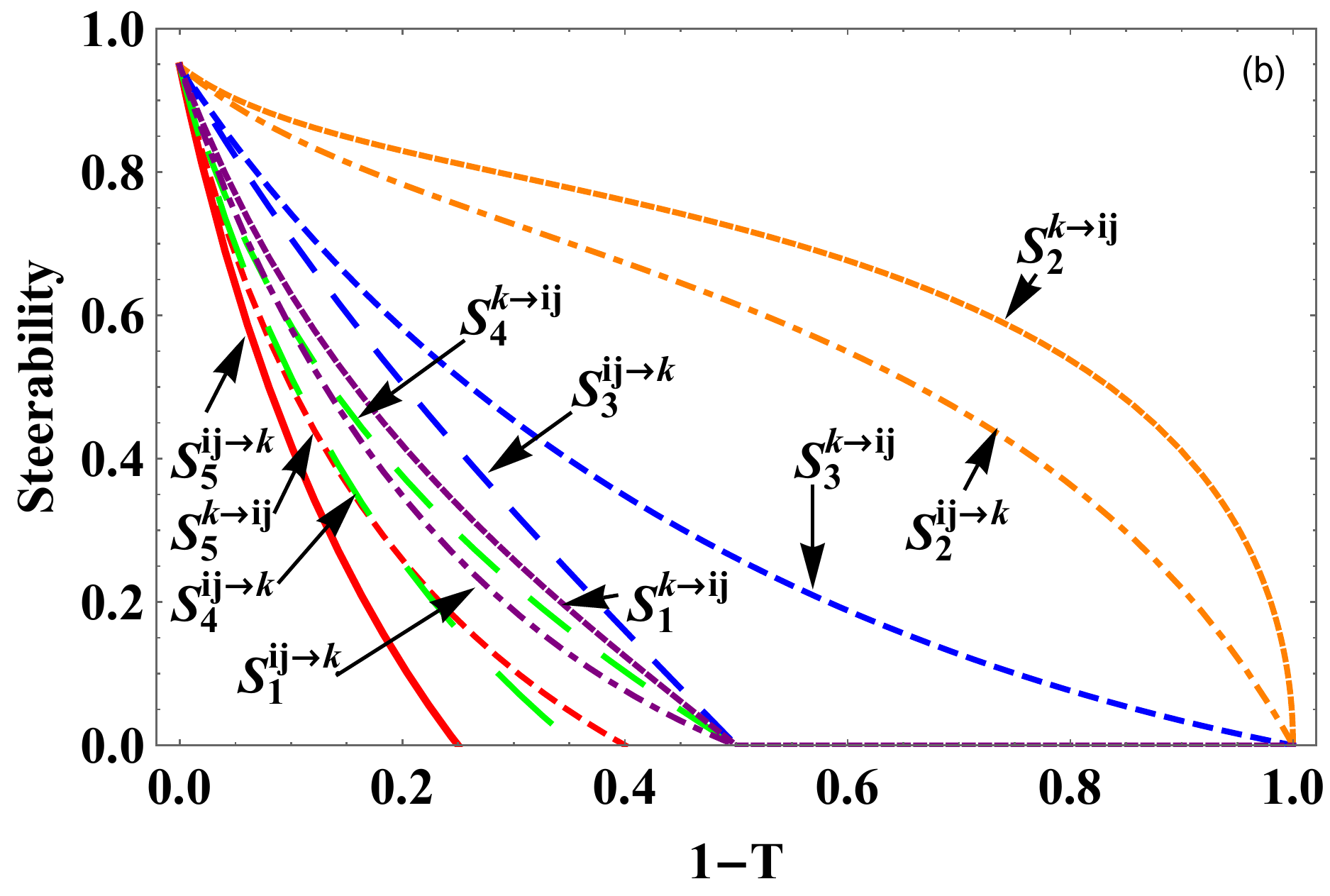}}
\caption{ Comparison of steering across all five loss scenarios for (a) $%
\protect\lambda=0.3$ and (b) $\protect\lambda=0.8$. The ranking Scenario 2 $%
> $ 3 $>$ 1 $>$ 4 $>$ 5 holds for both directions, confirming the critical
role of loss location. }
\label{fig:S_comparison}
\end{figure}

\subsection{Physical Significance of Loss Asymmetry}

The differences in correlation behavior under various loss configurations
reveal the sensitivity of quantum resources to environmental noise and its
intrinsic connection to system structure. We can rationalize the observed
robustness hierarchy by examining how each scenario disrupts the dominant
quadrature correlations. For instance, in Scenario 2, the loss is restricted
to one mode of the pair, while the other two modes remain lossless; the
unscathed mode preserves enough asymmetric information to sustain steering
in both directions. In contrast, Scenario 5 adds vacuum noise symmetrically
to all modes, rapidly washing out any directional bias.

These findings have immediate practical implications. For one-sided
device-independent quantum key distribution, the ability to maintain
steerability from a single trusted mode to an untrusted pair is crucial. Our
results show that this steerability can be maintained over a much wider
range of loss by minimizing the loss on the single trusted mode, even if the
other modes are subject to significant noise. This insight can guide the
design of fault-tolerant quantum networks where resources are allocated
asymmetrically based on channel quality.

\subsection{Correlation Hierarchy and Operational Distinction}

\begin{figure}[tbh]
\centering
\includegraphics[width=0.8\columnwidth]{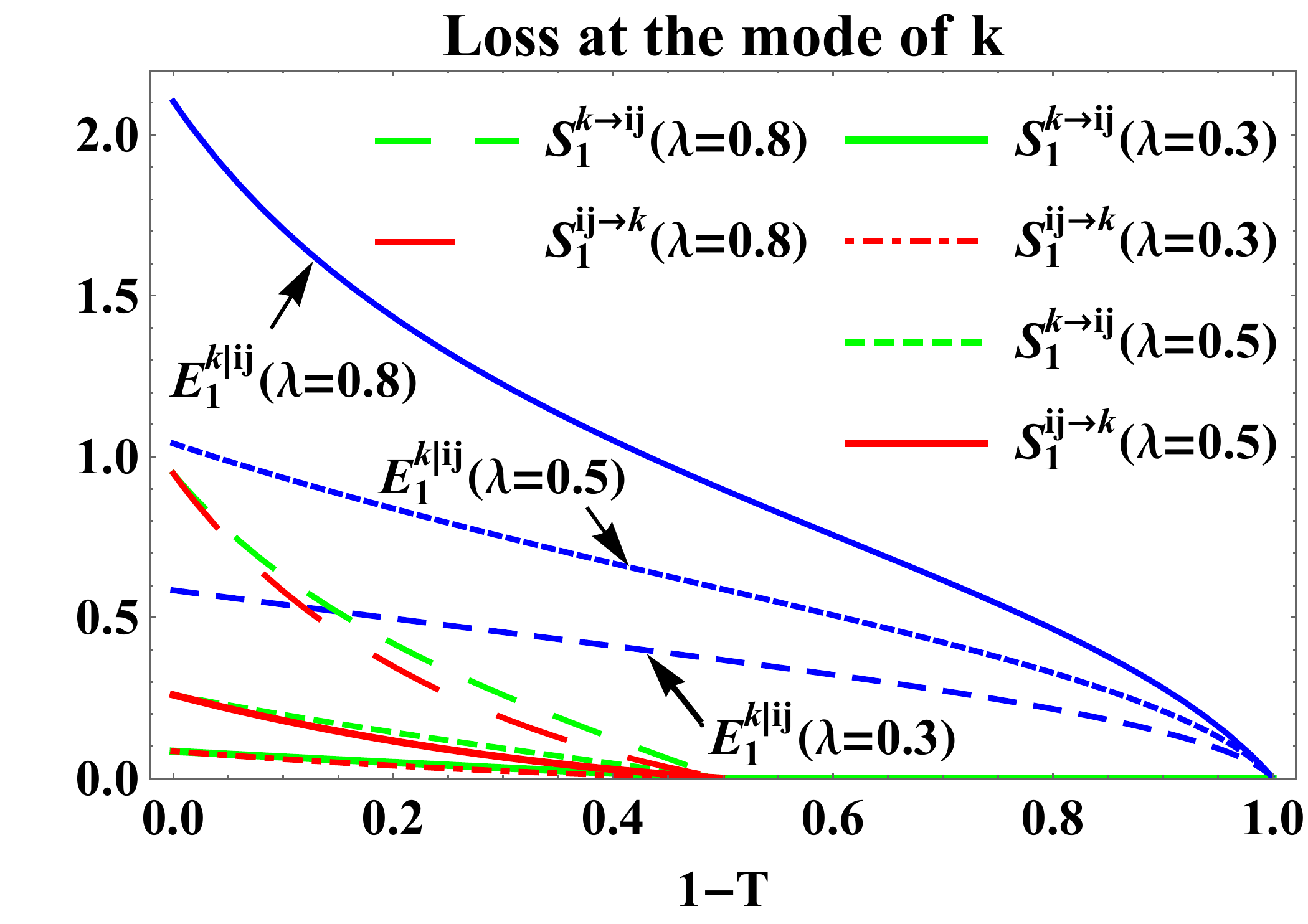}
\caption{ Correlation hierarchy in Scenario 1. As loss increases, steering
(dashed) vanishes before entanglement (solid), clearly exhibiting the strict
inclusion of steerable states within entangled states. }
\label{fig:hierarchy_1}
\end{figure}

\begin{figure}[tbh]
\centering
\includegraphics[width=0.8\columnwidth]{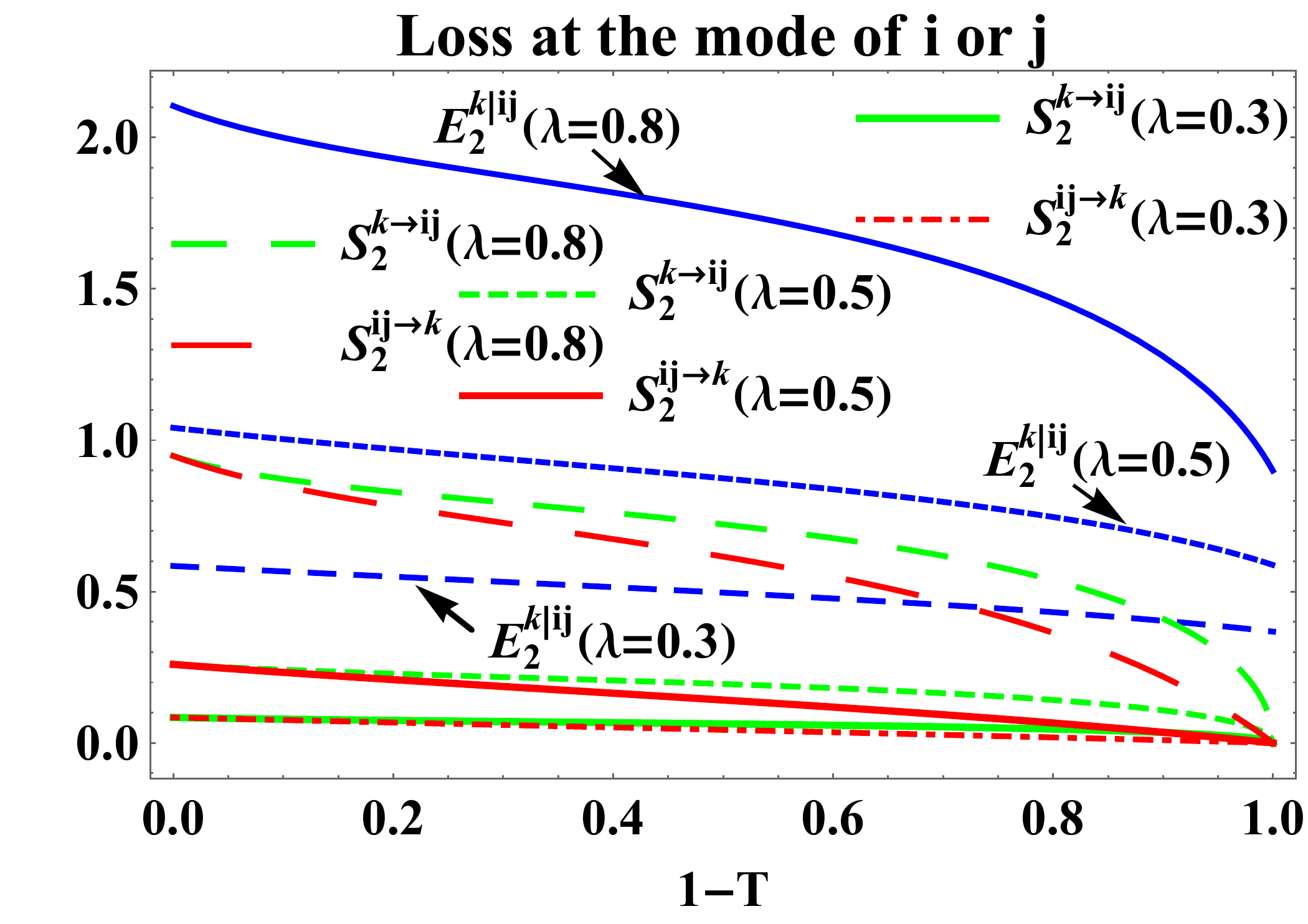}
\caption{ Correlation hierarchy in Scenario 2. Steering survives over the
entire range $T>0$, while entanglement is still more robust at high loss. }
\label{fig:hierarchy_2}
\end{figure}

\begin{figure}[tbh]
\centering
\includegraphics[width=0.8\columnwidth]{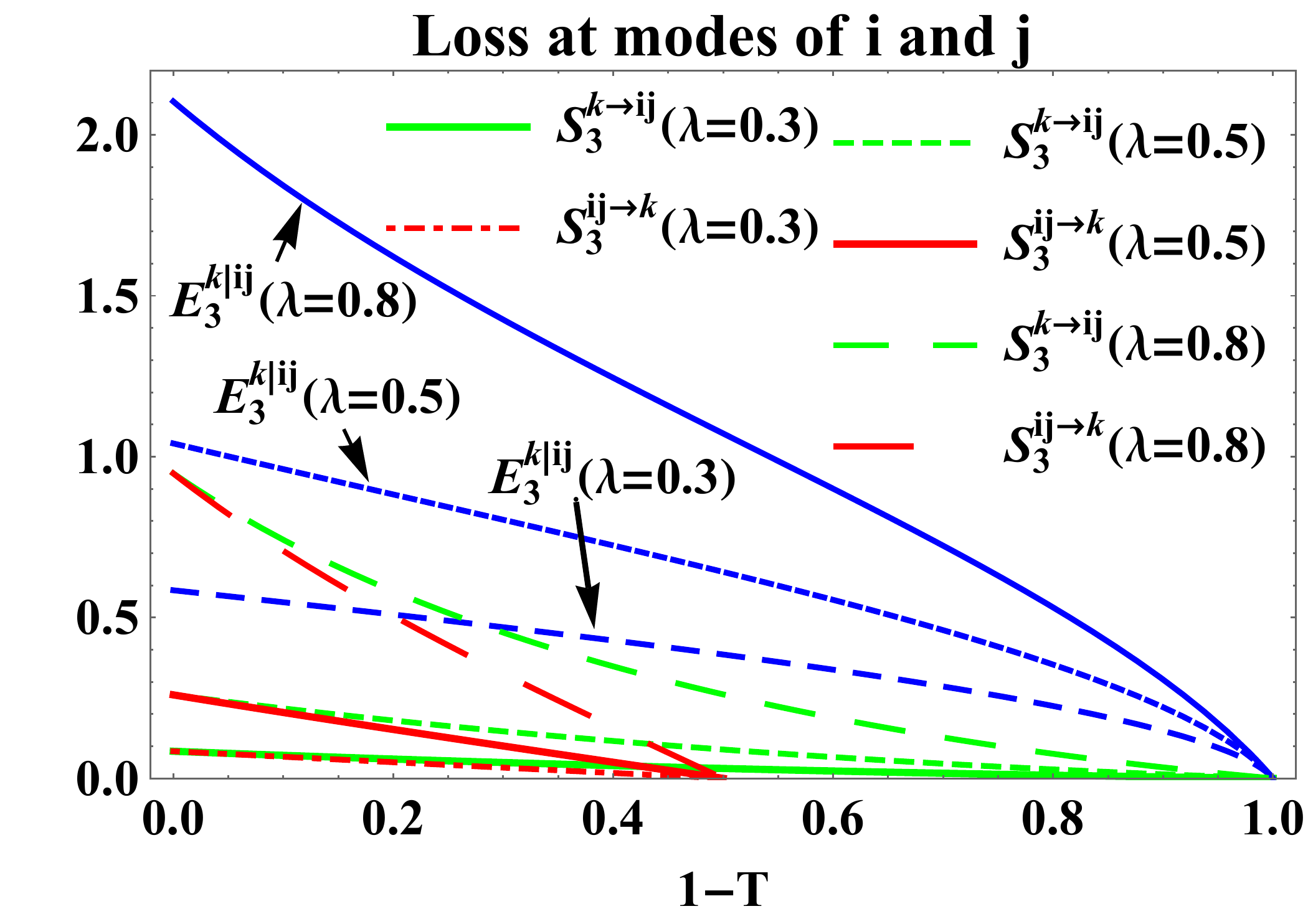}
\caption{Correlation hierarchy in Scenario 3.}
\label{fig:hierarchy_3}
\end{figure}

\begin{figure}[tbh]
\centering
\includegraphics[width=0.8\columnwidth]{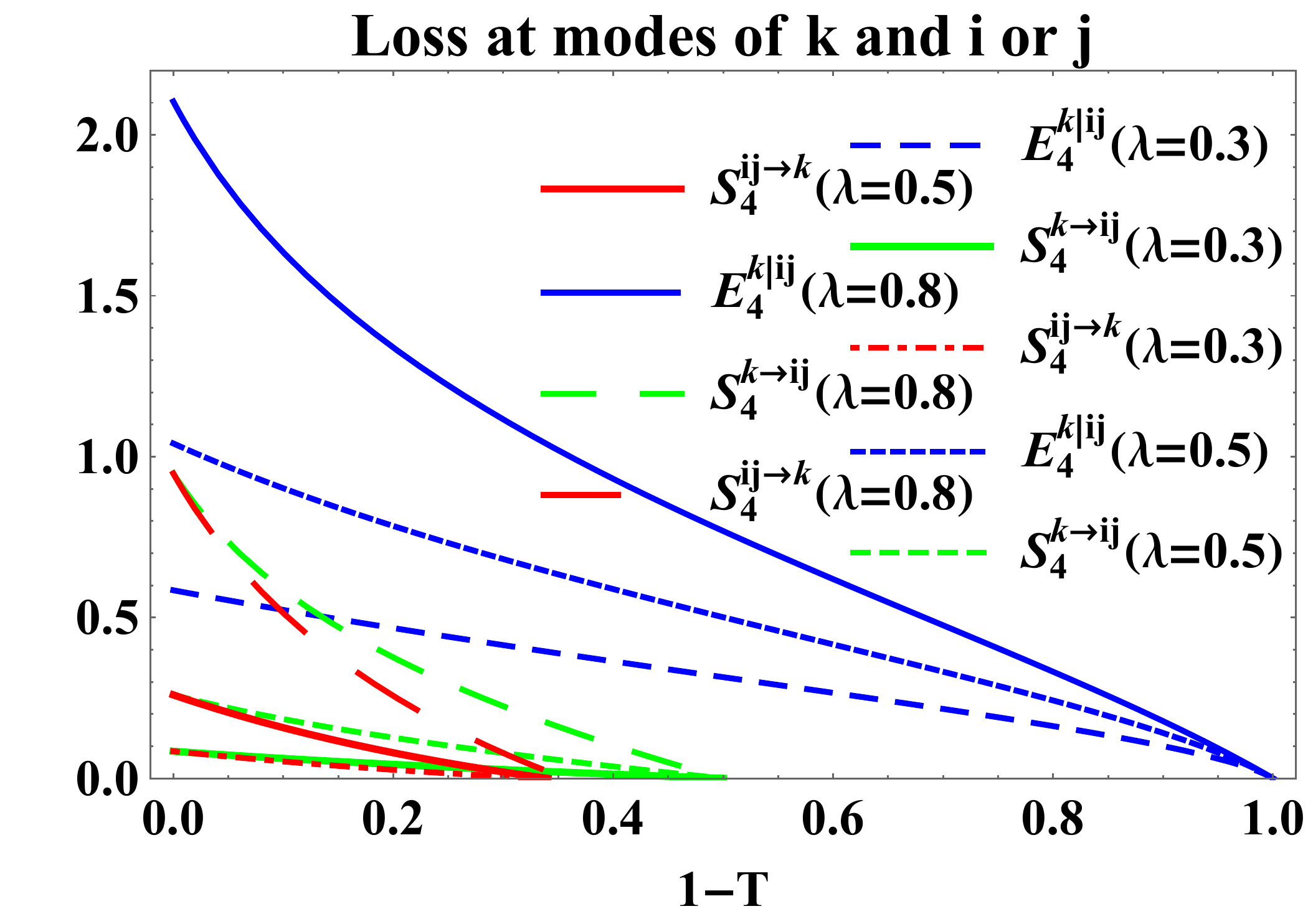}
\caption{Correlation hierarchy in Scenario 4.}
\label{fig:hierarchy_4}
\end{figure}

\begin{figure}[tbh]
\centering
\includegraphics[width=0.8\columnwidth]{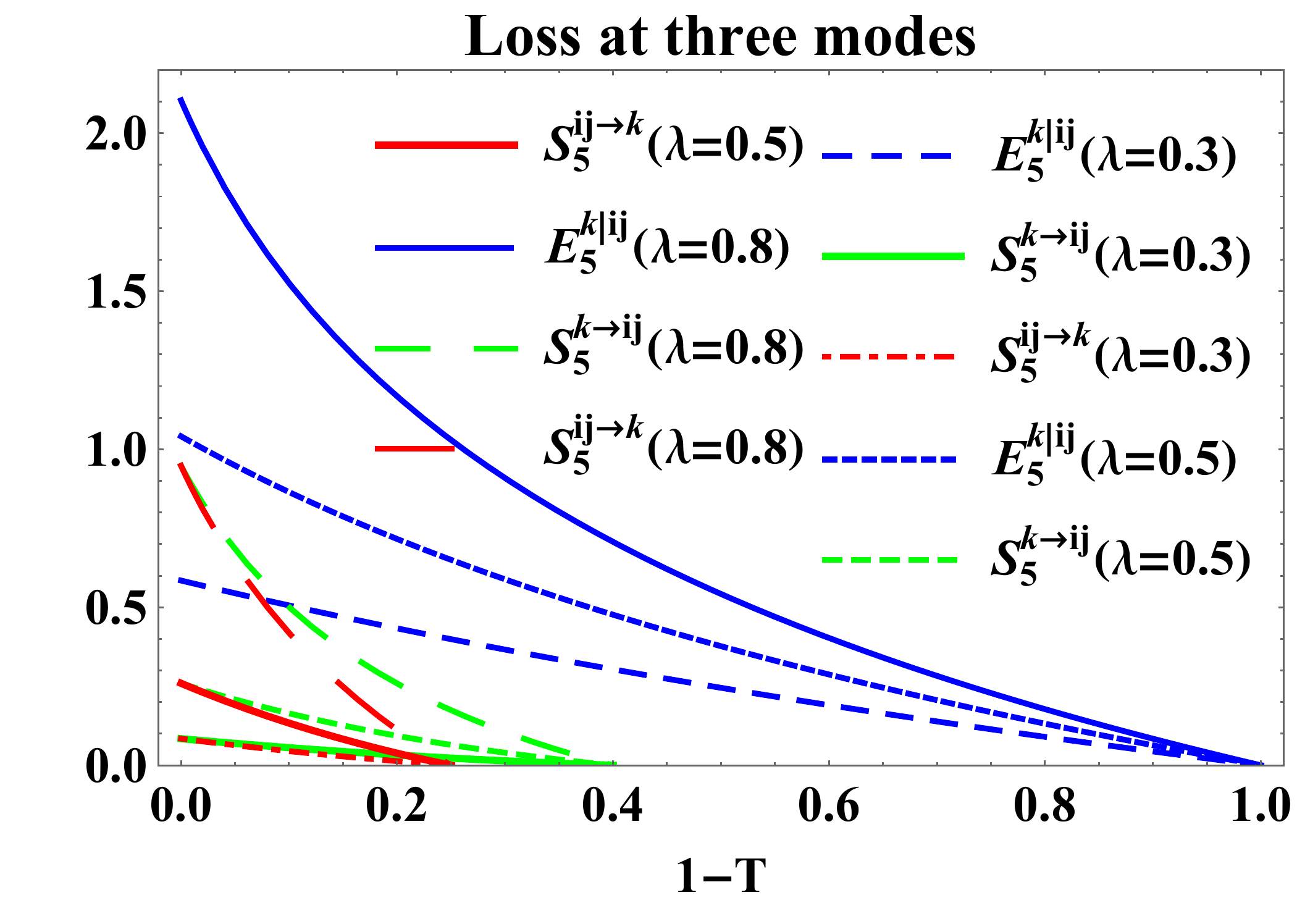}
\caption{ Correlation hierarchy in Scenario 5. Symmetric loss causes the
earliest disappearance of steering. }
\label{fig:hierarchy_5}
\end{figure}

The differential decay of entanglement and steering under loss provides a
clear operational demarcation within the quantum correlation hierarchy. As
shown in Figs.~\ref{fig:hierarchy_1}--\ref{fig:hierarchy_5}, for a fixed $%
\lambda$, as loss increases, the following sequence occurs:

\begin{enumerate}
\item \textbf{Region I (Low loss):} Both steering and entanglement are
present.

\item \textbf{Region II (Moderate loss):} Steering disappears (first $%
S^{ij\rightarrow k}$, then possibly $S^{k\rightarrow ij}$), while
entanglement remains.

\item \textbf{Region III (High loss):} Only entanglement remains until it
too eventually vanishes for extremely high loss.
\end{enumerate}

This confirms that the set of steerable states is a strict subset of the set
of entangled states for the tritter-generated state under loss. The monogamy
constraints [Eq.~(\ref{eq:monogamy})] continue to hold in all lossy
scenarios, as verified by our calculations.

\subsection{Revisiting the Hierarchy between Steering and Entanglement}

Our results clearly demonstrate the strict hierarchy of quantum
correlations: steering $\subset$ entanglement. This relationship is evident
not only in existence but also in robustness---entanglement persists at
higher noise levels. This difference stems from the distinct operational
definitions of the two resources: entanglement only requires non-classical
correlations, while steering further requires causal influence of one party
over the state of another \cite{Wiseman2007, Li2023}.

Our five-scenario analysis elevates this hierarchy from a qualitative
statement to a quantitative operational guide. By providing exact resilience
thresholds for each configuration, we enable experimental designers to
predict precisely how much loss can be tolerated before steering is lost,
and how to best allocate resources to maximize the survival of steerability.
This is a significant step beyond the abstract inclusion relation.

\section{Discussion and Outlook}

\label{sec:discussion}

\subsection{Complementary quantifiers: R\'{e}nyi-2 entropy}

To evaluate the performance of the Renyi-2 entropy in quantifying
entanglement and steering, we present Figure 14 based on the analysis
detailed in Appendix C. Figures (a) and (b) depict the tripartite Renyi-2
conditional mutual information and steering, respectively, as functions of
the reflectivity $1-T$. As shown in Figure (a), the conditional mutual
information remains positive across the entire range of total reflectivity
for all five scenarios. However, the relative magnitudes of the conditional
mutual information in these five scenarios differ from those obtained under
the positive partial transpose (PPT) criterion discussed earlier. As shown
in Figure (b), all steering curve values are greater than zero. Moreover,
the curves representing single-mode steering of the other two modes
consistently lie above those for the opposite direction. This observation
indicates that the Renyi-2 entropy is capable of revealing properties of
genuine tripartite steering. However, some steering curves do not decrease
monotonically with increasing reflectivity, which deviates from physically
expected behavior. Furthermore, a direct comparison of the two subplots does
not allow us to conclude that steering constitutes an entangled true subset.
In summary, the Renyi-2 entropy exhibits notable limitations in analyzing
entanglement and steering in continuous-variable states within multipartite
systems. A possible physical explanation is that the Renyi-2 entropy, as a
quantifier of entanglement and steering, essentially performs a simple
operation on the system's degree of mixing and thus fails to uncover deeper
correlations in quantum systems.

\begin{figure}[tbh]
\centering
\subfigure[]{\includegraphics[width=0.83\columnwidth]{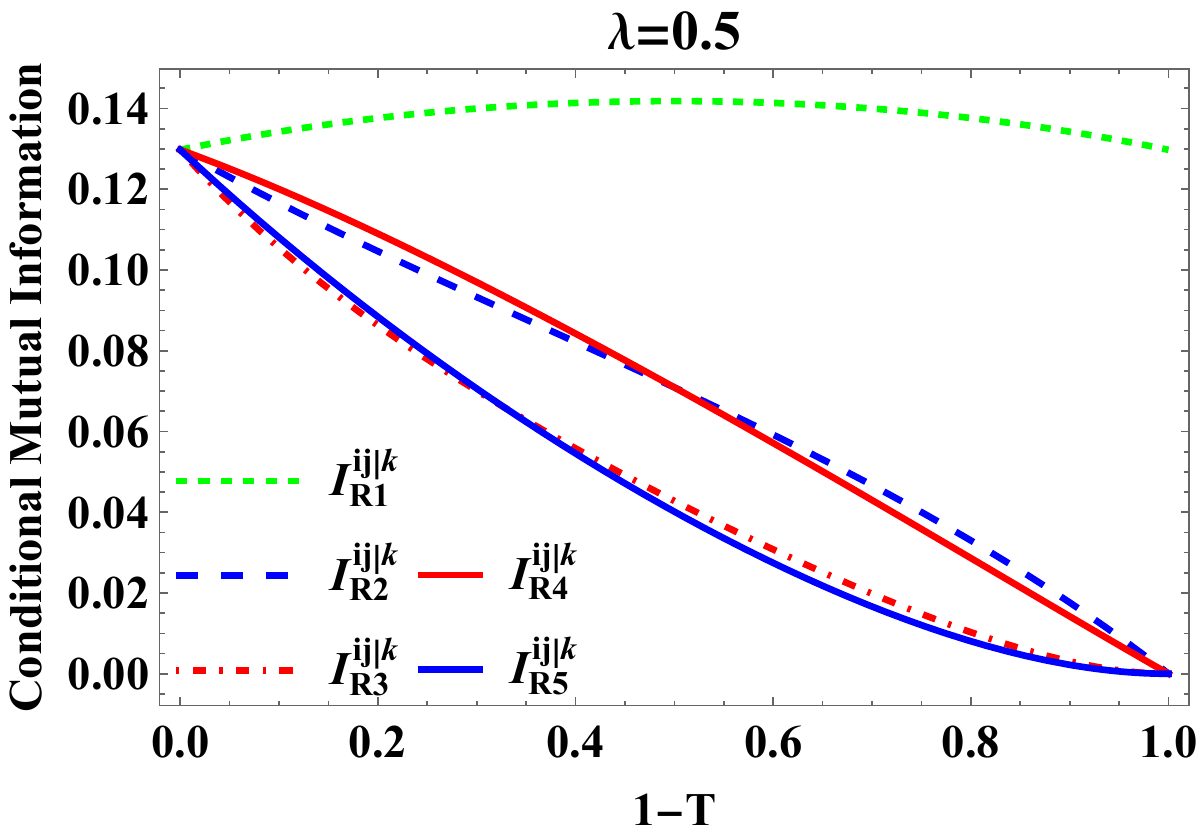}} \newline
\subfigure[]{\includegraphics[width=0.83\columnwidth]{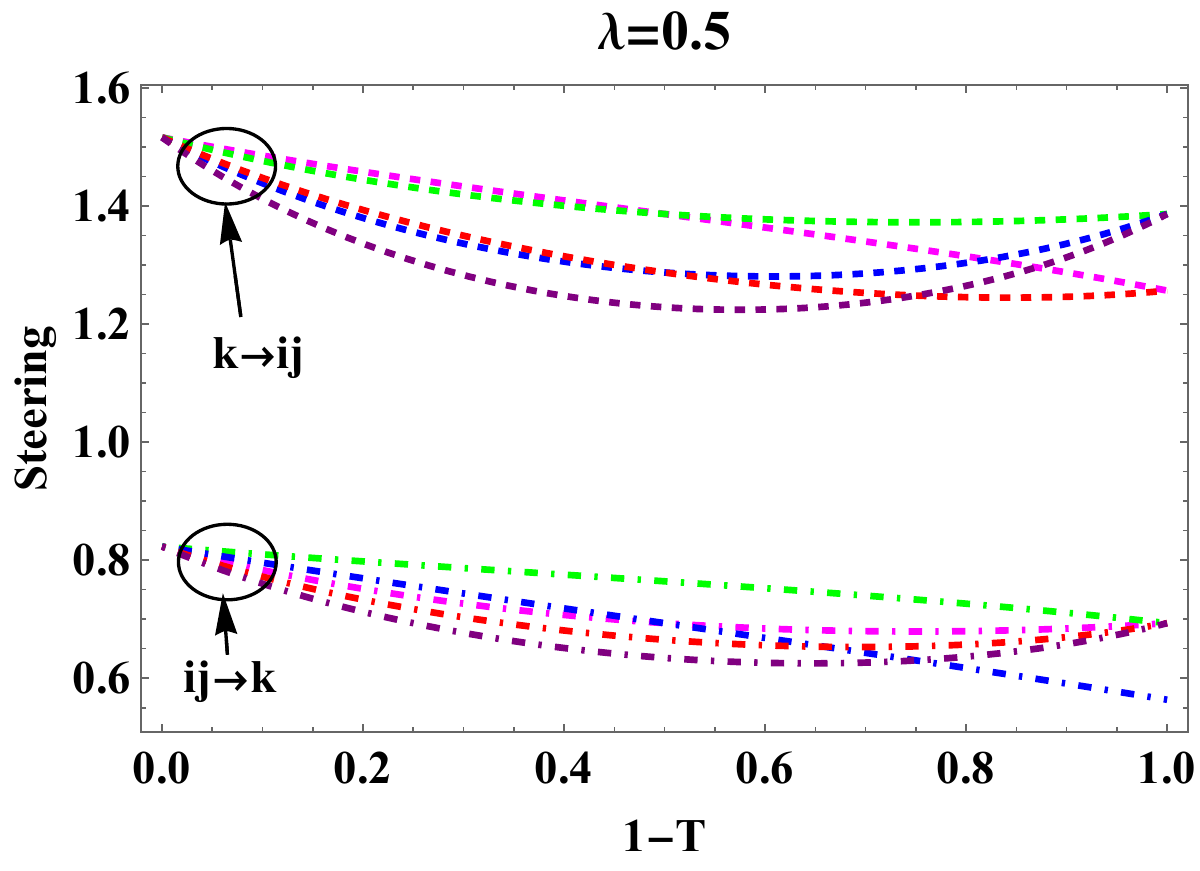}}
\caption{ Quantifying quantum correlation based on R\'{e}nyi-2 entropy($%
\protect\lambda=0.5$) (a) Conditional Mutual Information and (b)R\'{e}nyi-2
entropy Steering }
\label{fig14}
\end{figure}

\subsection{Generality of the findings}

Although our analysis focuses on five representative loss configurations,
the underlying physical mechanism suggests a broader universality. Steering
requires asymmetric correlation strength between parties, which
environmental mixing inherently degrades by adding uncorrelated vacuum
noise. Therefore, in any Gaussian system where loss couples locally to
individual modes, the volume of steerable states is expected to shrink
faster than that of entangled states. The partial protection
strategy---minimizing loss on the steering party---is also expected to be a
general design principle. Possible exceptions, such as non-Markovian
environments with backflow of information or engineered non-Gaussian
operations that can purify steering, present intriguing open questions.

\subsection{Scaling to $m>3$ modes}

The tritter is the $N=3$ member of a family of balanced multiport
interferometers. Our analysis can be extended to an $m$-mode tritter fed
with multiple TMSV sources and coherent states. Based on the three-mode
results, we expect that the resulting network will exhibit a complex
steering graph, where the directionality of steering between collective
modes can be tuned by local losses. In particular, the partial protection
strategy identified here (protecting a single mode to preserve steering
towards a larger group) is likely to generalize: in a star-network
configuration, minimizing the loss on the central node could maintain
steering to all other nodes even if they suffer significant attenuation. A
rigorous analysis of these $m$-mode scenarios is left for future work.

\section{Conclusion}

\label{sec:conclusion}

We have presented a comprehensive phase diagram of entanglement and EPR
steering in a three-mode Gaussian state generated by a tritter. By
systematically analyzing five physically distinct asymmetric loss
configurations and all bipartite partitions, we have mapped out the exact
resilience thresholds of both resources. Our results demonstrate that EPR
steering is not only a stricter resource than entanglement but also exhibits
a marked directional asymmetry under loss: steerability in the direction $%
k\to ij$ generally outlives $ij\to k$, and can be maintained over a much
wider range of loss by protecting a single mode. This tunable fragility
provides practical guidance for one-sided device-independent quantum
protocols. The observed hierarchy is robust, as confirmed by an independent
analysis using R\'{e}nyi-2 entanglement monotones. The underlying physical
mechanism---that steering relies on asymmetric correlations which local loss
degrades---suggests the generality of our findings for Gaussian systems.
Extensions to $m$-mode networks indicate that this partial protection
strategy could play a key role in the design of scalable quantum
communication architectures. Our work provides the analytical foundations
and practical insights needed to engineer noise-resilient quantum
correlations in realistic asymmetric environments.

\section*{Acknowledgments}

This work is supported by the National Natural Science Foundation of China
(Grants No. 12564049 and No. 12104195), the Jiangxi Provincial Natural
Science Foundation (Grants No. 20242BAB26009 and 20232BAB211033), the
Jiangxi Provincial Key Laboratory of Advanced Electronic Materials and
Devices (Grant No. 2024SSY03011), and the Jiangxi Civil-Military Integration
Research Institute (Grant No. 2024JXRH0Y07).

\appendix

\section{Derivation of the Covariance Matrix}

\label{app:cov}

Here we provide a self-contained derivation of the covariance matrix,
including the explicit tritter Bogoliubov transformation. The tritter
unitary matrix $U$ in Eq.~(\ref{eq:tritter_U}) is represented in the
symplectic picture by a $6\times6$ orthogonal symplectic matrix $S$.
Decomposing the complex amplitudes into quadratures, one obtains $S$ from $U$
via the standard mapping $\hat{x} = (\hat{a}+\hat{a}^\dagger)/\sqrt{2}$, $%
\hat{p} = (\hat{a}-\hat{a}^\dagger)/(i\sqrt{2})$. The full matrix $S$ can be
found in the Supplemental Material.

The characteristic function formalism provides an efficient way to compute
moments. The Wigner characteristic function for the output state $|\psi
_{in}\rangle $ is defined as $W_{in}(\zeta )=\langle \psi _{in}|\hat{D}%
(\zeta )|\psi _{in}\rangle $, where $\hat{D}(\zeta )=\exp (\zeta \widehat{A}%
^{+}-\zeta ^{\ast }\widehat{A})$ is the displacement operator, $\zeta
=\left( \lambda _{a},\lambda _{b},\lambda _{c}\right) ,\zeta ^{\ast }=\left(
\lambda _{a}^{\ast },\lambda _{b}^{\ast },\lambda _{c}^{\ast }\right) $, $%
\widehat{A}=\left( \widehat{a},\widehat{b},\widehat{c}\right) ^{T}$, and $%
\widehat{A}^{+}=\left( \widehat{a}^{+},\widehat{b}^{+},\widehat{c}%
^{+}\right) ^{T}$ . Based on Eq.(2), the results are as follows

\begin{equation}
W_{in}(\zeta )=\exp \left[ -\frac{\Gamma _{1}}{2}+\Gamma _{2}+\Gamma _{3}%
\right] ,
\end{equation}%
where $\Gamma _{1}=\frac{\left( 1+\lambda ^{2}\right) \left( \left\vert
\lambda _{a}\right\vert ^{2}+\left\vert \lambda _{b}\right\vert ^{2}\right)
}{\left( 1-\lambda ^{2}\right) }+\left\vert \lambda _{c}\right\vert ^{2}$, $%
\Gamma _{2}=\frac{\lambda \left( \lambda _{a}\lambda _{b}+\lambda _{a}^{\ast
}\lambda _{b}^{\ast }\right) }{\left( 1-\lambda ^{2}\right) }$ and $\Gamma
_{3}=\lambda _{c}\gamma ^{\ast }-\lambda _{c}^{\ast }\gamma $. As can be
seen from reference \cite{Chang2022}, the Wigner characteristic function
after the Tritter transformation is given by%
\begin{equation}
W_{out}(\zeta )=W_{in}(\zeta ^{^{\prime }}),
\end{equation}%
where $\zeta ^{^{\prime }}=\left( \lambda _{a}^{^{\prime }},\lambda
_{b}^{^{\prime }},\lambda _{c}^{^{\prime }}\right) $, $\zeta ^{^{\prime
}}=U^{-1}\zeta $.

According to the characteristic function theory, the relationship between
the normally ordered characteristic function and the Wigner characteristic
function is $N_{out}\left( \zeta \right) =e^{|\zeta |^{2}/2}W_{out}(\zeta )$
,to further obtain the expectation values of a generic operator given by
\begin{equation}
\left\langle
a^{+m_{1}}a^{m_{2}}b^{+n_{1}}b^{n_{2}}c^{+s_{1}}c^{s_{2}}\right\rangle
=\left. \widetilde{D}N_{out}\left( \zeta \right) \right\vert _{\lambda
_{j}=\lambda _{j}^{\ast }=0},
\end{equation}%
where $j=a,b,c$\ and $\widetilde{D}$ is a partial differential equations,\
its detailed description as follows

\begin{equation}
\widetilde{D}=\frac{\left( -\right) ^{m_{2}+n_{2}+s_{2}}\partial
^{m_{1}+m_{2}+n_{1}+n_{2}+s_{1}+s_{2}}}{\partial \lambda
_{a}^{^{m_{1}}}\partial \lambda _{a}^{^{\ast m_{2}}}\partial \lambda
_{b}^{^{n_{1}}}\partial \lambda _{b}^{^{\ast n_{2}}}\partial \lambda
_{c}^{^{s_{1}}}\partial \lambda _{c}^{^{\ast s_{2}}}}.
\end{equation}

Using Equation (A3) and (A4), we can obtain the following expression for the
expected value of operators.\

\begin{eqnarray}
\left\langle \hat{x}_{a}\right\rangle &=&\left\langle \hat{x}%
_{b}\right\rangle =\frac{3i\left( \gamma -\gamma ^{\ast }\right) -\sqrt{3}%
\left( \gamma +\gamma ^{\ast }\right) }{6\sqrt{2}}, \\
\left\langle \hat{p}_{a}\right\rangle &=&\left\langle \hat{p}%
_{b}\right\rangle =\frac{3\left( \gamma +\gamma ^{\ast }\right) +\sqrt{3}%
i\left( \gamma -\gamma ^{\ast }\right) }{6\sqrt{2}}, \\
\left\langle \hat{x}_{c}\right\rangle &=&\frac{\gamma +\gamma ^{\ast }}{%
\sqrt{6}},\left\langle \hat{p}_{c}\right\rangle =\frac{\gamma -\gamma ^{\ast
}}{i\sqrt{6}},
\end{eqnarray}

\begin{eqnarray}
\left\langle a^{+}a\right\rangle &=&\left\langle b^{+}b\right\rangle
=\left\langle c^{+}c\right\rangle =\frac{1}{3}\left( \left\vert \gamma
\right\vert ^{2}+\frac{2\lambda ^{2}}{1-\lambda ^{2}}\right) , \\
\left\langle a^{+2}\right\rangle &=&\left\langle b^{+2}\right\rangle =\frac{%
e^{i\frac{\pi }{3}}}{3}\left( e^{i\frac{\pi }{3}}\gamma ^{\ast 2}-\frac{%
2\lambda }{1-\lambda ^{2}}\right) , \\
\left\langle c^{+2}\right\rangle &=&\left\langle c^{2}\right\rangle ^{\ast }=%
\frac{e^{-i\frac{\pi }{3}}}{3}\left( e^{i\frac{\pi }{3}}\gamma ^{\ast 2}-%
\frac{2\lambda }{1-\lambda ^{2}}\right) , \\
\left\langle a^{2}\right\rangle &=&\left\langle b^{2}\right\rangle
=\left\langle a^{+2}\right\rangle ^{\ast }=\left\langle b^{+2}\right\rangle
^{\ast }, \\
\left\langle ab^{+}\right\rangle &=&\left\langle a^{+}b\right\rangle =\frac{1%
}{3}\left( \left\vert \gamma \right\vert ^{2}-\frac{\lambda ^{2}}{1-\lambda
^{2}}\right) , \\
\left\langle ac^{+}\right\rangle &=&\left\langle bc^{+}\right\rangle =\frac{%
e^{i\frac{2\pi }{3}}}{3}\left( \left\vert \gamma \right\vert ^{2}-\frac{%
\lambda ^{2}}{1-\lambda ^{2}}\right) , \\
\left\langle a^{+}c\right\rangle &=&\left\langle b^{+}c\right\rangle
=\left\langle ac^{+}\right\rangle ^{\ast }=\left\langle bc^{+}\right\rangle
^{\ast }, \\
\left\langle a^{+}b^{+}\right\rangle &=&\left\langle ab\right\rangle ^{\ast
}=\frac{e^{i\frac{2\pi }{3}}}{3}\left( \gamma ^{\ast 2}+\frac{e^{-i\frac{\pi
}{3}}\lambda }{1-\lambda ^{2}}\right) , \\
\left\langle a^{+}c^{+}\right\rangle &=&\left\langle b^{+}c^{+}\right\rangle
=-\frac{e^{i\frac{\pi }{3}}}{3}\left( \gamma ^{\ast 2}+\frac{e^{-i\frac{\pi
}{3}}\lambda }{1-\lambda ^{2}}\right) , \\
\left\langle ac\right\rangle &=&\left\langle bc\right\rangle =\left\langle
a^{+}c^{+}\right\rangle ^{\ast }=\left\langle b^{+}c^{+}\right\rangle ^{\ast
}.
\end{eqnarray}

From Eq. (A5)$\sim $(A17), the covariance matrix elements of the output
state are calculated as follows:

\begin{eqnarray}
C(\hat{x}_{i},\hat{x}_{i}) &=&\frac{3-2\lambda +\lambda ^{2}}{6(1-\lambda
^{2})}\quad (i=a,b,c), \\
C(\hat{p}_{i},\hat{p}_{i}) &=&\frac{3+2\lambda +\lambda ^{2}}{6(1-\lambda
^{2})}\quad (i=a,b,c), \\
C(\hat{x}_{i},\hat{p}_{i}) &=&C(\hat{p}_{i},\hat{x}_{i})=\frac{\sqrt{3}%
\lambda }{3(1-\lambda ^{2})}\quad (i=a,b), \\
C(\hat{x}_{c},\hat{p}_{c}) &=&C(\hat{p}_{c},\hat{x}_{c})=-\frac{\sqrt{3}%
\lambda }{3(1-\lambda ^{2})}, \\
C(\hat{x}_{a},\hat{x}_{b}) &=&C(\hat{x}_{b},\hat{x}_{a})=\frac{\lambda
(1-2\lambda )}{6(1-\lambda ^{2})}, \\
C(\hat{x}_{a},\hat{x}_{c}) &=&C(\hat{x}_{c},\hat{x}_{a})=\frac{\lambda
(\lambda -2)}{6(1-\lambda ^{2})}, \\
C(\hat{x}_{b},\hat{x}_{c}) &=&C(\hat{x}_{c},\hat{x}_{b})=\frac{\lambda
(\lambda -2)}{6(1-\lambda ^{2})}, \\
C(\hat{p}_{a},\hat{p}_{b}) &=&C(\hat{p}_{b},\hat{p}_{a})=-\frac{\lambda
(1+2\lambda )}{6(1-\lambda ^{2})}, \\
C(\hat{p}_{a},\hat{p}_{c}) &=&C(\hat{p}_{c},\hat{p}_{a})=\frac{\lambda
(\lambda +2)}{6(1-\lambda ^{2})}, \\
C(\hat{p}_{b},\hat{p}_{c}) &=&C(\hat{p}_{c},\hat{p}_{b})=\frac{\lambda
(\lambda +2)}{6(1-\lambda ^{2})}, \\
C(\hat{x}_{a},\hat{p}_{b}) &=&C(\hat{p}_{b},\hat{x}_{a})=-\frac{\sqrt{3}%
\lambda }{6(1-\lambda ^{2})}, \\
C(\hat{x}_{a},\hat{p}_{c}) &=&C(\hat{p}_{c},\hat{x}_{a})=\frac{\sqrt{3}%
\lambda ^{2}}{6(1-\lambda ^{2})}, \\
C(\hat{x}_{b},\hat{p}_{a}) &=&C(\hat{p}_{a},\hat{x}_{b})=-\frac{\sqrt{3}%
\lambda }{6(1-\lambda ^{2})},
\end{eqnarray}

\begin{align}
C(\hat{x}_{b},\hat{p}_{c})& =C(\hat{p}_{c},\hat{x}_{b})=\frac{\sqrt{3}%
\lambda ^{2}}{6(1-\lambda ^{2})}, \\
C(\hat{x}_{c},\hat{p}_{a})& =C(\hat{p}_{a},\hat{x}_{c})=-\frac{\sqrt{3}%
\lambda ^{2}}{6(1-\lambda ^{2})}, \\
C(\hat{x}_{c},\hat{p}_{b})& =C(\hat{p}_{b},\hat{x}_{c})=-\frac{\sqrt{3}%
\lambda ^{2}}{6(1-\lambda ^{2})}.
\end{align}

\section{Symmetric Eigenvalue}

In the analysis of quantum correlations based on the covariance matrix, the
symmetric eigenvalue table represents the core invariants of quantum
correlations (particularly entanglement and non-Gaussianity) in systems of
continuous variables: characterizing entanglement and steering (logarithmic
negativity); measures total correlation (R\'{e}nyi-2); and detects
non-Gaussianity and quantum phase transitions.In characterization of
entanglement, the symmetric eigenvalues can be obtained by taking the
modulus of the eigenvalues of the matrix $i\Omega \widetilde{\sigma }$. For
quantum steering of B, this is given by taking the absolute value of matrix $%
i\Omega V^{B|A}$. Where, $\widetilde{\sigma }$\ is the partial transpose of
the covariance matrix $\sigma $, and $V^{B|A}$ is the Schur complement of A.
Based on the above principles, the following table lists the required
symmetric eigenvalues for measuring entanglement\ and\ steering under five
different loss scenarios.It should be noted that the first digit in the
subscript represents the corresponding loss scenario, while the second digit
indicates the order of symmetric eigenvalues in that scenario.

\subsection{Symmetric Eigenvalue of entanglement}

\begin{eqnarray}
v_{11}^{ij|k} &=&v_{12}^{ij|k}=\frac{1}{2}, \\
v_{13}^{ij|k} &=&v_{14}^{ij|k}=\sqrt{\frac{9-\epsilon _{1}\lambda
^{2}+4\lambda \sqrt{18T+\epsilon _{2}\lambda ^{2}}}{36\left( 1-\lambda
^{2}\right) }}, \\
v_{15}^{ij|k} &=&v_{16}^{ij|k}=\sqrt{\frac{9-\epsilon _{1}\lambda
^{2}-4\lambda \sqrt{18T+\epsilon _{1}\lambda ^{2}}}{36\left( 1-\lambda
^{2}\right) }}, \\
v_{31}^{ij|k} &=&v_{32}^{ij|k}=\sqrt{\frac{1-\left( 1-2T\right) ^{2}\lambda
^{2}}{4\left( 1-\lambda ^{2}\right) }}, \\
v_{33}^{ij|k} &=&v_{34}^{ij|k}=\sqrt{\frac{9-\epsilon _{3}\lambda
^{2}+2\lambda \sqrt{72T+\epsilon _{4}\lambda ^{2}}}{36\left( 1-\lambda
^{2}\right) }}, \\
v_{35}^{ij|k} &=&v_{36}^{ij|k}=\sqrt{\frac{9-\epsilon _{3}\lambda
^{2}-2\lambda \sqrt{72T+\epsilon _{4}\lambda ^{2}}}{36\left( 1-\lambda
^{2}\right) }},
\end{eqnarray}

\begin{eqnarray}
v_{51}^{ij|k} &=&v_{52}^{ij|k}=\sqrt{\frac{1-\left( 1-2T\right) ^{2}\lambda
^{2}}{4\left( 1-\lambda ^{2}\right) }}, \\
v_{53}^{ij|k} &=&v_{54}^{ij|k}=\sqrt{\frac{9-\epsilon _{5}\lambda
^{2}+2T\lambda \sqrt{72+\epsilon _{6}\lambda ^{2}}}{36\left( 1-\lambda
^{2}\right) }}, \\
v_{55}^{ij|k} &=&v_{56}^{ij|k}=\sqrt{\frac{9-\epsilon _{5}\lambda
^{2}-2T\lambda \sqrt{72+\epsilon _{6}\lambda ^{2}}}{36\left( 1-\lambda
^{2}\right) }}.
\end{eqnarray}

It should be noted that the symmetric eigenvalues of entanglement\ in
Scenarios 2 and 4 are too complex to yield exact analytical expressions. In
each scenario, only the fifth and sixth symplectic eigenvalues can be less
than 1/2, i.e., $v_{l5}^{ij|k}=v_{l6}^{ij|k}<1/2\left( l=1,2,3,4,5\right) $,
which can be used to quantify entanglement.

\subsection{Symmetric Eigenvalue of steering}

\begin{eqnarray}
\overline{v}_{11}^{k|ij} &=&\overline{v}_{12}^{k|ij}=\sqrt{\frac{9-\left(
1-4T\right) ^{2}\lambda ^{2}}{4\left( 9-\lambda ^{2}\right) }}, \\
\overline{v}_{11}^{ij|k} &=&\overline{v}_{12}^{ij|k}=\frac{1}{2}, \\
\overline{v}_{13}^{ij|k} &=&\overline{v}_{14}^{ij|k}=\sqrt{\frac{9-\left(
1-4T\right) ^{2}\lambda ^{2}}{4\left[ 9-\left( 3-4T\right) ^{2}\lambda ^{2}%
\right] }}, \\
\overline{v}_{21}^{k|ij} &=&\overline{v}_{22}^{k|ij}=\sqrt{\frac{\left(
1-\lambda ^{2}\right) \left[ 9-\left( 1-4T\right) ^{2}\lambda ^{2}\right] }{%
4\left( 9-2\chi _{0}\lambda ^{2}+\lambda ^{4}\right) }}, \\
\overline{v}_{21}^{ij|k} &=&\overline{v}_{22}^{ij|k}=\frac{1}{2}\sqrt{1+%
\frac{\chi _{1}+4\lambda ^{2}\sqrt{T\chi _{2}}}{\left( 1-\lambda ^{2}\right)
\left( 9-\lambda ^{2}\right) }}, \\
\overline{v}_{23}^{ij|k} &=&\overline{v}_{24}^{ij|k}=\frac{1}{2}\sqrt{1+%
\frac{\chi _{1}-4\lambda ^{2}\sqrt{T\chi _{2}}}{\left( 1-\lambda ^{2}\right)
\left( 9-\lambda ^{2}\right) }}, \\
\overline{v}_{31}^{k|ij} &=&\overline{v}_{32}^{k|ij}=\frac{1}{2}\sqrt{\frac{%
9-\left( 1+2T\right) ^{2}\lambda ^{2}}{9-\left( 3-2T\right) ^{2}\lambda ^{2}}%
}, \\
\overline{v}_{31}^{ij|k} &=&\overline{v}_{32}^{ij|k}=\frac{1}{2}\sqrt{\frac{%
1-\left( 1-2T\right) ^{2}\lambda ^{2}}{1-\lambda ^{2}}}, \\
\overline{v}_{33}^{ij|k} &=&\overline{v}_{34}^{ij|k}=\frac{1}{2}\sqrt{\frac{%
9-\left( 1+2T\right) ^{2}\lambda ^{2}}{1-\lambda ^{2}}}, \\
\overline{v}_{41}^{k|ij} &=&\overline{v}_{42}^{k|ij}=\sqrt{\frac{%
9-2\vartheta _{r}\lambda ^{2}+\left( 1-4T^{2}\right) ^{2}\lambda ^{4}}{%
4\left( 9-2\chi _{0}\lambda ^{2}+\lambda ^{4}\right) }}, \\
\overline{v}_{41}^{ij|k} &=&\overline{v}_{42}^{ij|k}=\sqrt{\frac{9+\vartheta
_{0}\lambda ^{4}-2\lambda ^{2}\left( \vartheta _{1}-2\sqrt{\vartheta _{s}}%
\right) }{4\left( 1-\lambda ^{2}\right) \left[ 9-\left( 3-4T\right)
^{2}\lambda ^{2}\right] }}, \\
\overline{v}_{43}^{ij|k} &=&\overline{v}_{44}^{ij|k}=\sqrt{\frac{9+\vartheta
_{0}\lambda ^{4}-2\lambda ^{2}\left( \vartheta _{1}+2\sqrt{\vartheta _{s}}%
\right) }{4\left( 1-\lambda ^{2}\right) \left[ 9-\left( 3-4T\right)
^{2}\lambda ^{2}\right] }},
\end{eqnarray}

\begin{eqnarray}
\overline{v}_{51}^{k|ij} &=&\overline{v}_{52}^{k|ij}=\frac{1}{2}\sqrt{\frac{%
9-\left( 3-6T\right) ^{2}\lambda ^{2}}{9-\left( 3-2T\right) ^{2}\lambda ^{2}}%
}, \\
\overline{v}_{51}^{ij|k} &=&\overline{v}_{52}^{ij|k}=\frac{1}{2}\sqrt{\frac{%
1-\left( 1-2T\right) ^{2}\lambda ^{2}}{1-\lambda ^{2}}}, \\
\overline{v}_{53}^{ij|k} &=&\overline{v}_{54}^{ij|k}=\frac{1}{2}\sqrt{\frac{%
9-\left( 3-6T\right) ^{2}\lambda ^{2}}{9-\left( 3-4T\right) ^{2}\lambda ^{2}}%
},
\end{eqnarray}

where, $\chi _{0}=5-8T+8T^{2}$, $\vartheta _{r}=5-16T+20T^{2}$ and $%
\vartheta _{s}=\vartheta _{2}+\vartheta _{3}+\vartheta _{4}+\vartheta _{5}$.
For each scenario, both symmetric eigenvalues can be less than 1/2 in the
case of dual-mode steering single-mode, whereas only the two symmetric
eigenvalues are less than 1/2 in the case of the opposite direction
steering. So only expressions $\overline{v}_{l3}^{ij|k}$ and $\overline{v}%
_{l4}^{ij|k}$ can be used to quantify the steering.

\section{R\'{e}nyi-2 Entanglement and Steering Calculations}

\label{app:renyi}

For completeness, we provide here the basic overview of the Gaussian R\'{e}%
nyi-2 entanglement and steering.

The Gaussian R\'{e}nyi-2 entropy for a quantum states $\rho $ is defined as
\cite{Adesso2012}:
\begin{equation}
S_{2}\left( \rho \right) =\frac{1}{2}\ln \left( \det \sigma \right) ,
\end{equation}%
where $\sigma $\ is the covariance matrice of $\rho $. Using this formula,
we can obtain the following general expressions of R\'{e}nyi-2 entropy

\begin{eqnarray}
S_{2}\left( \rho _{i}\right) &\thickapprox &\frac{1}{2}\ln \left\{ \frac{%
0.028\left[ 9-\left( 3-4T_{i}\right) ^{2}\lambda ^{2}\right] }{1-\lambda ^{2}%
}\right\} , \\
S_{2}\left( \rho _{jk}\right) &=&\frac{1}{2}\ln \left\{ \frac{9-2\left(
9+\digamma _{1}\right) \lambda ^{2}+\digamma _{2}^{2}\lambda ^{4}}{144\left(
1-\lambda ^{2}\right) ^{2}}\right\} , \\
S_{2}\left( \rho _{ijk}\right) &=&\frac{1}{2}\ln \left\{ \frac{9-2\digamma
_{3}\lambda ^{2}+\digamma _{4}^{2}\lambda ^{4}}{576\left( 1-\lambda
^{2}\right) ^{2}}\right\} ,
\end{eqnarray}%
where $\digamma _{1}=8\left( T_{j}^{2}+T_{k}^{2}\right)
+4T_{j}T_{k}-12\left( T_{j}+T_{k}\right) $, $\digamma
_{2}=3+4T_{j}T_{k}-4\left( T_{j}+T_{k}\right) $,$\digamma _{3}=9+8\left(
T_{i}^{2}+T_{j}^{2}+T_{k}^{2}\right) +4\left(
T_{i}T_{j}+T_{i}T_{k}+T_{j}T_{k}\right) -12\left( T_{i}+T_{j}+T_{k}\right) $,%
$\digamma _{4}=3+4\left( T_{i}T_{j}+T_{j}T_{k}+T_{i}T_{k}\right) -4\left(
T_{i}+T_{j}+T_{k}\right) $.

Using the GR-2 mutual information $I_{2}\left( \rho _{A},\rho _{B}\right) $,
we can measure the total correlation between two subsystems, A and B, of a
quantum system (including classical correlation and quantum entanglement).
According to references\cite{Adesso2012}, the mutual information of
quantum states $\rho _{AB}$\ can be expressed as

\begin{equation}
I_{2}\left( A,B\right) =S_{2}\left( \rho _{A}\right) +S_{2}\left( \rho
_{B}\right) -S_{2}\left( \rho _{AB}\right) ,
\end{equation}%
For the tripartite states $\rho _{ABC}$, the Renyi-2 conditional mutual
information\cite{Jonah2023} is defined as

\begin{equation}
I_{2}\left( A,B|C\right) =S_{2}\left( \rho _{AC}\right) +S_{2}\left( \rho
_{BC}\right) -S_{2}\left( \rho _{ABC}\right) -S_{2}\left( \rho _{C}\right) ,
\end{equation}

Moreover, the R\'{e}nyi-2 steering correlation between $A$ and $B$ \cite%
{Huang2024} is:%
\begin{eqnarray}
S_{R}^{A\rightarrow B} &=&S_{2}\left( \rho _{A}\right) -S_{2}\left( \rho
_{AB}\right) =\ln \sqrt{\frac{\det \sigma _{A}}{\det \sigma _{AB}}}, \\
S_{R}^{B\rightarrow A} &=&S_{2}\left( \rho _{B}\right) -S_{2}\left( \rho
_{AB}\right) =\ln \sqrt{\frac{\det \sigma _{B}}{\det \sigma _{AB}}},
\end{eqnarray}%
where $\sigma _{AB}$ is the covariance matrice of quantum states $\rho _{AB}$%
.It should be noted that the above conclusions also apply to some
non-Gaussian states\cite{Zhang2022}.

\end{document}